\documentclass{aa}  
\usepackage{graphicx}
\usepackage{float}
\usepackage{rotating}
\usepackage[section]{placeins}
\usepackage{amsmath}
\usepackage{mathastext}
\newcommand{\angstrom}{\mbox{\normalfont\AA}}
\usepackage[svgnames]{xcolor}
\usepackage{url}
\usepackage{csvsimple}
\usepackage{pgfplotstable}
\usepackage{longtable}
\usepackage[varg]{txfonts}
\usepackage[version=4]{mhchem}
\usepackage{isotope}
\usepackage{units} 
\usepackage{comment}
\usepackage{threeparttablex}
\usepackage{geometry}

\usepackage{lscape}
\usepackage{booktabs}
\usepackage{array}
\pgfplotsset{compat=1.17}
\usepackage{verbatim}
\usepackage{multicol} 
\usepackage{multirow}
\usepackage{colortbl}

\makeatletter
\renewcommand*\aa@pageof{, page \thepage{} of \pageref*{LastPage}}

\makeatletter
\newcommand\footnoteref[1]{\protected@xdef\@thefnmark{\ref{#1}}\@footnotemark}
\makeatother

\usepackage{hyperref} 
\hypersetup{breaklinks=true, colorlinks=true, linkcolor=blue, citecolor=blue, urlcolor=blue}

\usepackage{natbib,twoopt} 
\bibpunct{(}{)}{;}{a}{}{,}  
\defcitealias{brauner23}{Paper I}

\begin{document}

   \title{Unveiling the chemical fingerprint of phosphorus-rich stars}
   \subtitle{II. Heavy-element abundances from UVES/VLT spectra\thanks{Line list, modifications to the line list, systematic uncertainties, and abundances of the light elements are only available in electronic form at the CDS via anonymous ftp to cdsarc.u-strasbg.fr (130.79.128.5) or via \url{http://cdsweb.u-strasbg.fr/cgi-bin/qcat?J/A+A/}.}}
   
   \author{Maren Brauner\inst{1}\fnmsep\inst{2}
          \and
          Marco Pignatari\inst{3}\fnmsep\inst{4}\fnmsep\inst{5}\fnmsep\inst{6}\fnmsep\inst{7}
          \and
          Thomas Masseron\inst{1}\fnmsep\inst{2}
          \and
          D. A. García-Hernández\inst{1}\fnmsep\inst{2} 
          \and 
          Maria Lugaro\inst{3}\fnmsep\inst{4}\fnmsep\inst{8}\fnmsep\inst{9}
          }

   \institute{Instituto de Astrofísica de Canarias, C/Via Láctea s/n, E-38205 La Laguna,      Tenerife, Spain\\
              \email{maren.brauner@iac.es}
        \and
             Departamento de Astrofísica, Universidad de La Laguna, E-38206 La Laguna, Tenerife, Spain
        \and
            Konkoly Observatory, Research Centre for Astronomy and Earth Sciences, HUN-REN, H-1121 Budapest, Konkoly Thege M. út 15-17, Hungary
        \and
            CSFK, MTA Centre of Excellence, Budapest, Konkoly Thege Miklós út 15-17, H-1121, Hungary
        \and
            E. A. Milne Centre for Astrophysics, University of Hull, Hull, HU6 7RX, UK
        \and
            Joint Institute for Nuclear Astrophysics - Center for the Evolution of the Elements
        \and
            The NuGrid Collaboration, http://www.nugridstars.org
        \and
            ELTE Eötvös Loránd University, Institute of Physics and Astronomy, Budapest 1117, Pázmány Péter sétány 1/A, Hungary
        \and
            School of Physics and Astronomy, Monash University, VIC 3800, Australia
            \\
             }

   \date{Received; accepted }

 
  \abstract
   {The atmospheres of phosphorus-rich (P-rich) stars have been shown to contain between 10 and 100 times more P than our Sun. Given its crucial role as an essential element for life, it is especially necessary to uncover the origin of P-rich stars to gain insights into the still unknown nucleosynthetic formation pathways of P in our Galaxy.}
   {Our objective is to obtain the extensive chemical abundance inventory of four P-rich stars, covering a large range of heavy (Z > 30) elements. This characterization will serve as a milestone for the nuclear astrophysics community to uncover the processes that form the unique chemical fingerprint of P-rich stars.}
   {We performed a detailed 1D local thermodynamic equilibrium abundance analysis on the optical UVES spectra of four P-rich stars. The abundance measurements, complemented with upper-limit estimates, included 48 light and heavy elements. Our focus lay on the neutron-capture elements (Z $> 30$), in particular, on the elements between Sr and Ba, as well as on Pb, as they provide valuable constraints to nucleosynthesis calculations. In past works, we showed that the heavy-element observations from the first P-rich stars are not compatible with either classical s-process or r-process abundance patterns. In this work, we compare the obtained abundances with three different nucleosynthetic scenarios: a single i-process, a double i-process, and a combination of s- and i-processes.
   }
   {We have performed the most extensive abundance analysis of P-rich stars to date, including the elements between Sr and Ba, such as Ag, which are rarely measured in any type of stars. We also estimated constraining upper limits for Cd I, In I, and Sn I. We found overabundances with respect to solar in the s-process peak elements, accompanied by an extremely high Ba abundance and slight enhancements in some elements between Rb and Sn. No global solution explaining all four stars could be found for the nucleosynthetic origin of the pattern. The model that produces the least number of discrepancies in three of the four stars is a combination of s- and i-processes, but the current lack of extensive multidimensional hydrodynamic simulations to follow the occurrence of the i-process in different types of stars makes this scenario highly uncertain. 
   }
   {}
   
   \keywords{stars: chemically peculiar -- 
             stars: abundances --
             nuclear reactions, nucleosynthesis, abundances --
            }

   \maketitle
%
%
\section{Introduction\label{Introduction}}
\noindent
Since their discovery by \citet{masseron20a}, phosphorus-rich (P-rich) stars are a challenging enigma for stellar nucleosynthesis modelers. Not only is the pure existence of stars with abnormally high amounts of P a mystery to theoreticians, the underlying origin of the P content in our Galaxy also remains unclear because Galactic chemical evolution (GCE) models systematically underestimate the existing amount and predict up to three times less P than is observed \citep{cescutti12,caffau11, ritter:18, prantzos:18, kobayashi:20}. \\
\indent The chemically peculiar P-rich stars were first reported by \citet{masseron20a}, who analyzed the near-infrared (NIR) H-band spectra of 15 stars extracted from the APOGEE\footnote{Apache Point Observatory Galactic Evolution Experiment}-2 data release 14 (DR14; \citealp{abolfathi18APOGEEDR14}). In addition to high P abundances [P/Fe]\footnote{\label{Bracket}$\left[\frac{X}{Fe}\right]=\log\left(\frac{N_X}{N_H}\right)_{*}-\log\left(\frac{N_X}{N_H}\right)_{\odot}-\left[\frac{Fe}{H}\right]$\\ with~$\left[\frac{Fe}{H}\right]=\log\left(\frac{N_{Fe}}{N_H}\right)_{*}-\log\left(\frac{N_{Fe}}{N_H}\right)_{\odot}$ and $N_X$ the number of atoms of element X.}$\gtrsim$\unit[1.0]{dex}, these metal-poor ([Fe/H] $\approx$ \unit[$-$1.0]{dex}) giants also show significant enhancements in O, Mg, Al, Si, and Ce.
The results from \citet{masseron20a} were confirmed by \citet{brauner23} (hereafter \citetalias{brauner23}), who enlarged the P-rich sample to a total of 78 stars with similar abundance patterns and also showed correlations between the enhanced elements. \\
\indent We know that the stellar mass of P-rich giants is about \unit[1]{M$_\odot$}. It is therefore evident that they are not able to synthesize these high amounts of P themselves. For this reason, the search for the origin of their abundances is focused on the progenitor that polluted the interstellar medium (ISM) from which the P-rich stars were born that we observe today. Several exotic pollution sources, binary mass transfer or the possibility of an extragalactic origin, have already been ruled out as a cause for the unusual abundance pattern (for a detailed review of the considered scenarios, see \citet{masseron20a,masseron20b} and \citetalias{brauner23}). In short, the progenitor of the P-rich stars is yet to be identified. \\
\indent In order to extend the abundance data, \citet{masseron20b} collected optical spectra of the two brightest P-rich stars using the Fibre-fed Echelle Spectrograph (FIES) at the \unit[2.5]{m} Nordic Optical Telescope (NOT). These spectra provide access to a larger number of light and heavy elements. Special attention was given to representative elements such as Ba, Eu, and Pb, which were compared with the prevailing nucleosynthetic processes. \\
\indent We recall that elements heavier than Fe are produced by two main neutron-capture processes that are defined by the neutron density: the r(rapid)- and s(slow)-process. The s- and r-process occur at low ($\lesssim\unit[10^{10}]{cm^{-3}}$) or high ($\gtrsim\unit[10^{20}]{cm^{-3}}$) neutron density, respectively. It is generally accepted that the s-process takes place during hydrostatic burning in massive and asymptotic giant branch (AGB) stars (e.g., \citealp{karakas14}). For the r-process, rare types of supernovae (SNe) and compact object mergers are currently debated as the main sites (e.g., \citealp{kajino19}). The discovery of metal-poor and post-AGB stars with a chemical pattern that is not explained by the s- or r-process \citep{barbuy97,jonsell2006} raised the need for an intermediate neutron-capture process, the i-process \citep{cowan1977}. This process is thought to result from the ingestion of protons in a convective He-burning region. The specific stellar sites in which the i-process may occur are not yet discovered, but based on current stellar simulations, most of the types of evolved stars (in particular at low metallicity) are possible hosts, such as low-metallicity AGB stars (e.g., \citealp{cristallo:09, hampel19,karinkuzhi20}), super-AGB stars \citep{doherty15, jones:16}, massive stars \citep[][]{roederer:16, clarkson:18,banerjee:18}, and accreting white dwarfs \citep{denissenkov17} 
\citep[see][for a recent review]{lugaro23}. \\
\indent To constrain the nucleosynthetic process that produced the pattern observed in the P-rich stars, \citet{masseron20b} compared the obtained heavy-element abundance pattern with the pattern of four metal-poor stars that represent the three neutron-capture processes. The overabundances that they found, namely in the first- (Sr, Y, and Zr) and the second-peak s-process elements (Ba, La, Ce, and Nd), along with the negative [Rb/Sr], point to a low neutron density s-process. Nevertheless, the nature of the s-process must be different from that in AGB stars because the P-rich stars show a higher [Ba/La] ratio and lower Eu and Pb abundances than AGB stars. Although AGB and super-AGB stars have already been ruled out by \citet{masseron20a}, who argued that P production in these stars is thought to be negligible \citep{Karakas10,doherty15}, P production through the s-process by neutron capture on \ce{^30Si} is still an option \citep{masseron20b}. Hence, the still-unidentified P-rich star progenitor might constitute a new stellar site for the production of heavy neutron-capture elements through some variation of the s-process. This has important implications for the chemical evolution of our Galaxy, the cosmic origin of the elements heavier than Fe, and possibly, owing to the production of P, even for the origin of life on Earth. \\
\indent Similar to \citetalias{brauner23}, this work is also dedicated to the chemical fingerprint of P-rich stars. Here, our main focus lay on the abundances of heavy elements as obtained from high-resolution spectra in the optical region, taken with the Ultraviolet and Visual Echelle Spectrograph (UVES; \citealp{dekker2000}) at the Very Large Telescope (VLT). The four selected targets are confirmed P-rich stars, and determining their heavy-element abundance pattern will help us to distinguish the nucleosynthetic process that occurred in their progenitor. We obtained the abundances of elements with Z > 30, including critical elements that are rarely probed, for instance, those from Nb to Sn. When no measurement was possible, we estimated an upper limit on the abundance. These are mostly constraining and likewise useful for the following analysis because the elements between Sr and Ba provide valuable information for nucleosynthetic calculations. The obtained patterns are compared to models of the s- and i-process, more precisely: models produced by one single i-process exposure, a double i-process exposure, and a combination of s- and i-process exposures. The i-process was previously ruled out by \citet{masseron20b} based on the observation of only two targets. Adding four more targets will clarify whether the i-process can definitely be excluded as a production mechanism. By following the suggestion of \citet{masseron20b} to observe more P-rich stars in the optical, we were also able to determine a more reliable abundance of the third peak s-process element Pb because \citet{masseron20b} was only able to measure the Pb abundance in one star. We also report the abundance results of all accessible light species with Z $\leq$ 30, which we determined to define the stellar atmosphere as well as possible. With a total number of 48 elemental abundances, including some upper limits, this is the first extensive abundance analysis of P-rich stars to date.


\section{Targets and observations\label{ObsData}}

\begin{table*}
\caption{Observation log and measured barycentric radial velocities $V_{\gamma}$.}           
\label{table:obslog}      
\centering                          
\begin{tabular}{c c c c c c c c c}        
\hline\hline
    Star & R.A. [J2000] & Decl. [J2000] & B & V & S/N & S/N & V$_{\gamma}$ & Obs. Date \\ 
    & [deg] & [deg] & [mag] & [mag] & \unit[3800]{\angstrom} & \unit[5000]{\angstrom} & [km/s] & \\ \hline
    2M00044180 & 00:04:41.806 & -00:05:55.320 & 13.59 & 12.51 & 172 & 201 & -57.71$\pm$0.15 & July 19/20, 2021 \\
    \dots & \dots & \dots & \dots & \dots & \dots & \dots & -57.24$\pm$0.30 & Jul. 20/21, 2021 \\
    \dots & \dots & \dots & \dots & \dots & \dots & \dots & -56.77$\pm$0.23 & Jul. 21/22, 2021 \\
    \dots & \dots & \dots & \dots & \dots & \dots & \dots & -54.76$\pm$0.09 & Aug. 9/10, 2021 \\
    2M13472354 & 13:47:23.539 & +22:10:56.389 & 12.99 & 12.35 & 147 & 299 & 36.20$\pm$0.09 & Jun. 28/29, 2021 \\
    \dots & \dots & \dots & \dots & \dots & \dots & \dots & 34.96$\pm$0.32 & Jul. 6/7, 2021 \\
    \dots & \dots & \dots & \dots & \dots & \dots & \dots & 35.14$\pm$0.23 & Jul. 8/9, 2021 \\
    \dots & \dots & \dots & \dots & \dots & \dots & \dots & 34.81$\pm$0.02 & Jul. 11/12, 2021 \\
    2M18453994 & 18:45:39.945 & -30:10:46.560 & 14.64 & 12.89 & 30 & 117 & 167.19$\pm$0.08 & Jun. 3/4, 2021 \\
    \dots & \dots & \dots & \dots & \dots & \dots & \dots & 166.34$\pm$0.41 & Jul. 6/7, 2021 \\
    \dots & \dots & \dots & \dots & \dots & \dots & \dots & 166.98$\pm$0.22 & Jul. 15/16, 2021 \\
    \dots & \dots & \dots & \dots & \dots & \dots & \dots & 168.32$\pm$0.26 & Aug. 6/7, 2021 \\
    2M22480199 & 22:48:01.993 & +14:11:32.949 & 13.13 & 11.89 & 116 & 153 & -227.14$\pm$0.21 & Jul. 6/7, 2021 \\
    \dots & \dots & \dots & \dots & \dots & \dots & \dots & -220.99$\pm$0.18 & Jul. 15/16, 2021 \\
    \dots & \dots & \dots & \dots & \dots & \dots & \dots & -218.50$\pm$0.25 & Jul. 19/20, 2021 \\
\hline
\end{tabular}
\tablefoot{Magnitudes B and V are taken from \citet{zacharias12} for 2M00044180 and 2M18453994, and \citet{hog00} for 2M13472354 and 2M22480199. The S/N is given per pixel.}
\end{table*}

We further investigated two targets in the optical that are the two brightest stars from \citet{masseron20a}. They are observable with the VLT and are 2M00044180-0005553 and 2M13472354+2210562. Two bright Si-rich stars from \citet{fernandeztrincado20}, 2M18453994-3010465 and 2M22480199+1411329, were also selected after a possible correlation between the P and Si abundances was demonstrated \citep{masseron20a}. Their status as promising P-rich candidates was confirmed prior to the submission of the proposal. Both stars were part of the previous study from \citetalias{brauner23}, who reconfirmed that they belonged to the group of P-rich stars. \\
\indent The observations of the four targets were carried out between June 3 and August 10, 2021, with UVES \citep{dekker2000} at the \unit[8.2]{m} diameter Unit Telescope 2 (UT2; Kueyen) of the VLT, mounted at the European Southern Observatory in Paranal, Chile. To cover the full wavelength range \unit[3300--9400]{\angstrom}, the two standard settings that employ the dichroic filters were employed, except for 2M13472354. This resulted in a spectral range from \unit[3300--6640]{\angstrom} for this star. The total exposure time per target of $\text{\unit[36,000]{s}}$ was distributed with the aim to reach a signal-to-noise ratio (S/N) > 100 over the whole spectral range. The slit width was fixed at \unit[0.7]{$''$}, resulting in a resolving power of R$\sim$62,000. For the reduction and extraction procedure, the ESO in-house pipeline was used. \\
\indent The observation log and the barycentric radial velocities $V_{\gamma}$ are given in Table \ref{table:obslog}. To perform the Doppler correction and determine $V_{\gamma}$, we used the \verb|IRAF|\footnote{IRAF is distributed by the National Optical Astronomy Observatory, which is operated by the Association of Universities for Research in Astronomy, Inc., under cooperative agreement with the United States National Science
Foundation.} (\citealp{tody93}, version 2.16.1) tasks \textit{fxcor} and \textit{dopcor}. The Doppler correction was obtained by cross-correlating the single exposures with a measured spectrum of Arcturus \citep{hinkle00arcturus}. For consistency reasons, we used the correlation between real measurements to determine the $V_{\gamma}$. We confirmed the results by repeating the cross-correlation with a synthetic spectrum of the Sun and a synthetic spectrum with the stellar parameters of the targets. In both cases, the range of the template matching comprises the full range of the target spectra without the edges, the CCD gaps of the spectrograph, and the ranges that are most affected by telluric contamination. The values given in Table \ref{table:obslog} represent the average over the $V_{\gamma}$ values obtained from the exposures on the same night, including the spectra of lower quality, and the errors are the corresponding standard deviations. We combined the individual spectra with the \verb|IRAF| task \textit{scombine}. We only combined spectra with the highest quality in order to achieve an S/N as high as possible in the blue and red parts. Estimates of the S/N are given in Table \ref{table:obslog}. We normalized the spectra within the abundance calculation procedure described in Sect. \ref{calculation}.

\section{Spectral analysis\label{AbuAnalysis}}

\begin{table*}
\caption{Atmospheric parameters of the targets.}           
\label{table:targetinfo}      
\centering                          
\begin{tabular}{c c c c c c c c c}        
\hline\hline
    Star & T$_{eff}$  & $\log$ g  & $\mu_t$ & $\text{[M/H]}$ & $\text{[Fe/H]}$\tablefootmark{a} & $\text{[Fe/H]}$\tablefootmark{b} & $\text{[P/Fe]}$\tablefootmark{b}  \\ 
    & [K] & [dex] & [km $s^{-1}$] & [dex] & [dex] & [dex] & [dex] \\ \hline
    2M00044180-0005553 & 4643 & 1.74 & 1.45 & -1.13 & -1.13 $\pm$ 0.15 & -1.03 $\pm$ 0.05 & 1.17 $\pm$ 0.13 \\
    2M13472354+2210562 & 5418 & 2.30 & 1.32 & -1.30 & -1.34 $\pm$ 0.18 & -1.21 $\pm$ 0.09 & 1.43 $\pm$ 0.01 \\
    2M18453994-3010465 & 3977 & 0.33 & 2.22 & -1.24 & -1.62 $\pm$ 0.21 & -1.22 $\pm$ 0.09 & 1.28 $\pm$ 0.13 \\
    2M22480199+1411329 & 4975 & 2.60 & 1.29 & -1.20 & -1.20 $\pm$ 0.15 & -1.10 $\pm$ 0.08 & 1.24 $\pm$ 0.13 \\
\hline
\end{tabular}
\tablefoot{Effective temperature T$_{eff}$, surface gravity $\log$ g, and overall metallicity $\text{[M/H]}$ are those provided by ASPCAP \citep{garciaperez16ASPCAP} using the APOGEE-2 spectra of DR17 \citep{abdurrouf22APOGEEDR17} for 2M00044180 and 2M22480199, and DR16 \citep{ahumada19APOGEEDR16} for 2M13472354 and 2M18453994. The microturbulence $\mu_t$ is obtained from the \textit{Gaia}-ESO microturbulence calibration \citep{smiljanic14,masseron19}. The errors of [Fe/H] and [P/Fe] correspond to the standard deviation of the line-by-line abundances. \\
\tablefoottext{a}{This work.}
\tablefoottext{b}{Result from \citetalias{brauner23}.}
}
\end{table*}

\subsection{Line lists and preparation\label{LL_prep}}
The atomic data for our abundance analysis consists of the most recent line list from the Vienna Atomic Line Database (VALD3\footnote{\url{http://vald.astro.uu.se/~vald/php/vald.php}}; \citealp{piskunow95,ryabchikova15}), including hyperfine structure (HFS) splitting \citep{pakhomov19}. We downloaded the blue part (\unit[3900--4500]{\angstrom}) of the line list on May 10, 2023, and the red part (\unit[4500--9500]{\angstrom}) on May 25, 2023. To account for the molecular bands of CH, $\ce{C2}$, and CN, we employed the molecular line lists from \citet{masseron14CH}, \citet{yurchenko18}, and \citet{sneden14CN}, respectively. \\
\indent In an attempt to have as many lines as possible at our disposal for determining the abundances, we performed a thorough search for atomic lines with the help of the solar spectrum. More precisely, we produced a synthesis of the solar spectrum using the 1D local thermal equilibrium (LTE) turbospectrum radiative transfer code v19.1.3 \citep{plez12}, and the MARCS model atmosphere of the Sun \citep{gustafsson08} with parameters T$_{eff}$=\unit[5777]{K}, $\log$ g = \unit[4.44]{dex}, and a fixed microturbulence velocity $\mu_t$ of \unit[1]{km/s}.
In the first step, we used the standard chemical composition from \citet{Grevesse2007}. In a second step, we produced new synthetic solar spectra, enhancing the absolute solar abundance of each element considered in this work by \unit[$+$1.00]{dex}. By comparing the synthetic solar spectra obtained in the first and second step, we selected the lines that were most prominent, and therefore, owing to their sensitivity to abundance alterations, were most suitable for the analysis. This search was performed over the whole wavelength range in which the P-rich stars were observed (\unit[3300--9400]{\angstrom}). The final selection of lines that we used for the abundance determination of the heavy elements (Z $> 30$), alongside the corresponding excitation potential (E.P.) and transition probabilities log(gf) taken from the VALD are available at the CDS. \\
\indent In many cases, such as some transition metals or rare earths (see Table \ref{table:abuheavy}), we had a large number of spectral lines after the search described above. While in this scenario we trusted in the derived average of the line-by-line abundances, other species only exhibited a small number of lines and therefore required a more careful treatment. The abundance determination of the elements we are most interested in for this study, for example, Ru to Sn or Pb, is most challenging. When only a small number of lines was available, we searched for possible blends in these critical lines to adjust the elements well that caused the blending. We also identified discrepancies in terms of line strengths between the synthetic solar spectrum and an observed spectrum \citep{neckel99solar}. This was done for the lines of the element in question and for nearby blends. When a line was not well reproduced by the synthetic spectrum produced with the VALD data, we adjusted the oscillator strength of the line to improve the agreement. 
An example for such a modification is the Ni I line at \unit[4093.036]{\angstrom}, which is a blend of the Hf I line at \unit[4093.150]{\angstrom}. We found that in the solar and target synthetic spectra, this line was synthesized too strongly, and we therefore slightly decreased the VALD log(gf)=--0.925 to log(gf)=--1.125. This resulted in a better match between the synthetic and observed spectra. \\
\indent The atmospheric parameters of the four targets we used as input for the following abundance calculations are summarized in Table \ref{table:targetinfo}. The effective temperature T$_{eff}$, surface gravity $\log$ g, and overall metallicity $\text{[M/H]}$ are taken from the APOGEE Stellar Parameters and Chemical Abundance Pipeline (ASPCAP; \citealp{garciaperez16ASPCAP}), the tool that provides parameters derived from APOGEE-2 spectra, in the case of 2M00044180 and 2M22480199 from DR17 \citep{abdurrouf22APOGEEDR17}, and for 2M13472354 and 2M18453994 from DR16 \citep{ahumada19APOGEEDR16}, because ASPCAP does not provide parameters for these two targets in DR17\footnote{Both targets are flagged in DR17 (\texttt{STAR\_BAD}), indicating issues when calculating the parameters, for example, a large number of bad pixels or that the solution lies at grid edge in T$_{eff}$ or $\log$ g.}. The microturbulence velocity $\mu_t$ is based on the \textit{Gaia}-ESO microturbulence calibration \citep{smiljanic14,masseron19}. In Table \ref{table:targetinfo} we also provide the P abundance ratio $\text{[P/Fe]}$ and metallicity $\text{[Fe/H]}$ derived in \citetalias{brauner23} alongside the metallicity $\text{[Fe/H]}$ obtained in this work. The discrepancy between $\text{[Fe/H]}$ from our work and the literature values is discussed in Sect. \ref{Fe}.

\begin{table*}
\caption{Abundances [X/Fe] of the heavy (Z > 30) elements together with the line-to-line standard deviation $\sigma_{\text{[X/Fe]}}$ and the number N of absorption lines.}             
\label{table:abuheavy}      
\centering                          
\begin{tabular}{c | c c c | c c c | c c c | c c c}        
\hline\hline                 
 & \multicolumn{3}{c}{2M00044180-0005553} & \multicolumn{3}{c}{2M13472354+2210562} & \multicolumn{3}{c}{2M18453994-3010465} & \multicolumn{3}{c}{2M22480199+1411329} \\    
                      
 & [X/Fe] & $\sigma_{\text{std}}$ & N & [X/Fe] & $\sigma_{\text{std}}$ & N & [X/Fe] & $\sigma_{\text{std}}$ & N & [X/Fe] & $\sigma_{\text{std}}$ & N \\
 \hline
 Ga I & <0.29 & \dots & \dots & -0.05 & 0.15 & 1 & <1.24 & \dots & \dots & 0.00 & 0.15 & 1 \\
 Ge I & 0.17 & 0.15 & 1 & <0.34 & \dots & \dots & 0.68 & 0.15 & 1 & 0.29	& 0.15 & 1 \\
 Rb I & 0.49 & 0.15 & 1 & \dots & \dots & \dots & 0.86 & 0.10 & 2 & 0.45 & 0.15 & 1 \\
 Sr I+II & 0.62 & 0.12 & 3 & 0.69 & 0.10 & 2 & 0.99 & 0.18 & 2 & 0.51 & 0.10 & 3 \\
 Sr I & 0.65 & 0.15 & 2 & \dots & \dots & \dots & 0.99 & 0.18 & 2 & 0.61 & 0.15 & 1 \\
 Sr II & 0.56 & 0.15 & 1 & 0.69 & 0.10 & 2 & \dots & \dots & \dots & 0.45 & 0.10 & 2 \\
 Y I+II & 0.36 & 0.13 & 21 & 0.54 & 0.18 & 19 & 0.56 & 0.21 & 13 & 0.45 & 0.12 & 24 \\
 Y I & 0.22 & 0.09 & 3 & \dots & \dots & \dots & 0.48 & 0.24 & 4 & 0.26 & 0.15 & 1 \\
 Y II & 0.39 & 0.12 & 18 & 0.54 & 0.18 & 19 & 0.59 & 0.20 & 9 & 0.46 & 0.11 & 23 \\
 Zr I+II & 0.63 & 0.13 & 29 & 0.66 & 0.20 & 30 & 1.00 & 0.12 & 11 & 0.62 & 0.11 & 27 \\
 Zr I & 0.63 & 0.11 & 17 & \dots & \dots & \dots & 0.98 & 0.13 & 9 & 0.58 & 0.08 & 11 \\
 Zr II & 0.64 & 0.14 & 12 & 0.66 & 0.20 & 30 & 1.08 & 0.10 & 2 & 0.65 & 0.12 & 16 \\
 Nb I+II & 0.60 & 0.12 & 4 & <0.92 & \dots & \dots & 0.83 & 0.18 & 5 & <1.28 & \dots & \dots \\
 Nb I & 0.62 & 0.14 & 3 & <0.92 & \dots & \dots & 0.83 & 0.18 & 5 & <1.28 & \dots & \dots \\
 Nb II & 0.53 &  0.15 & 1 & \dots & \dots & \dots & \dots & \dots & \dots & \dots & \dots & \dots \\
 Mo I & 0.79 & 0.05 & 5 & 0.47 & 0.15 & 1 & 0.95 & 0.13 & 7 & 0.64 & 0.05 & 4 \\
 Ru I & 0.46 & 0.09 & 6 & 0.18 & 0.10 & 2 & 0.82 & 0.10 & 2 & 0.48 & 0.05 & 4 \\
 Rh I & 0.28 & 0.10 & 2 & <0.34 & \dots & \dots & <0.50 & \dots & \dots & 0.45 & 0.18 & 3 \\
 Pd I & 0.29 & 0.15 & 1 & <0.57 & \dots & \dots & <0.46 & \dots & \dots & 0.42 & 0.14 & 2 \\
 Ag I & 0.09 & 0.15 & 1 & 0.31 & 0.15 & 1 & <0.18 & \dots & \dots & 0.25 & 0.15 & 1 \\
 Cd I & <0.51 & \dots & \dots & <0.82 & \dots & \dots & <1.85 & \dots & \dots & <0.83 & \dots & \dots \\
 In I & <-0.22 & \dots & \dots & <0.64 & \dots & \dots & <0.08 & \dots & \dots & <-0.30 & \dots & \dots \\
 Sn I & <0.63 & \dots & \dots & <0.59 & \dots & \dots & <0.68 & \dots & \dots & <0.01 & \dots & \dots \\
 Cs I & <1.56 & \dots & \dots & \dots & \dots & \dots & <1.55 & \dots & \dots & <1.50 & \dots & \dots \\
 Ba II & 1.07 & 0.10 & 4 & 1.61 & 0.20 & 4 & 1.85 & 0.08 & 4 & 0.88 & 0.11 & 4 \\
 La II & 0.54 & 0.05 & 21 & 0.56 & 0.09 & 19 & 0.83 & 0.07 & 11 & 0.60 & 0.07 & 20 \\
 Ce II & 0.37 & 0.07 & 40 & 0.42 & 0.05 & 37 & 0.64 & 0.15 & 19 & 0.45 & 0.05 & 26 \\
 Pr II & 0.35 & 0.14 & 16 & 0.25 & 0.05 & 3 & 0.68 & 0.08 & 11 & 0.49 & 0.08 & 12 \\
 Nd II & 0.37 & 0.08 & 35 & 0.29 & 0.05 & 29 & 0.63 & 0.07 & 26 & 0.46 & 0.05 & 45 \\
 Sm II & 0.23 & 0.06 & 15 & 0.16 & 0.05 & 13 & 0.42 & 0.06 & 9 & 0.37 & 0.05 & 18 \\
 Eu II & 0.28 & 0.05 & 5 & 0.04 & 0.07 & 4 & 0.59 & 0.05 & 6 & 0.40 & 0.08 & 6 \\
 Dy II & 0.28 & 0.05 & 6 & 0.04 & 0.08 & 10 & 0.49 & 0.13 & 3 & 0.36 & 0.10 & 12 \\
 Lu II & <0.57 & \dots & \dots & <1.03 & \dots & \dots & <1.06 & \dots & \dots & <0.64 & \dots & \dots \\
 Hf II & 0.29 & 0.15 & 1 & 0.47 & 0.15 & 1 & 0.34 & 0.15 & 1 & 0.44 & 0.15 & 1 \\
 Pb I & 0.42 & 0.10 & 2 & 0.47 & 0.15 & 1 & 0.42 & 0.14 & 2 & 0.53 & 0.10 & 2 \\ 
\hline
\end{tabular}
\tablefoot{The averages over all lines independent of the ionization state (indicated with "I+II") are given because we use the values in the pattern plots. The iron abundance used for the [X/Fe] ratio for the neutral and ionized lines was obtained from the Fe I lines (see Sect. \ref{Fe}).}
\end{table*}

\subsection{Abundance calculations\label{calculation}}
The procedure of abundance calculation in the current work is largely the same as in \citetalias{brauner23}. The elemental abundances were derived by employing the Brussels Automatic Code for Characterizing High accUracy Spectra (BACCHUS, version $\sim$v67; \citealp{masseron2016}), a tool that generates 1D LTE synthetic spectra by means of the code turbospectrum, v19.1.3 \citep{plez12}, taking as ingredients the line lists described in the previous section and MARCS model atmosphere grids \citep{gustafsson08, jonsson20}, interpolated at the corresponding parameters of the targets. As a result, BACCHUS provides the abundances on a line-by-line basis obtained from comparison of the synthetic and observed spectra. \\
\indent We briefly recall the five different comparison methods of BACCHUS that we used to derive the abundance value of a line (for a more detailed description, we refer to \citetalias{brauner23}). We also emphasize that different from \citetalias{brauner23}, we did not rely on only one method. Instead, we chose a method for each line case individually. Our general approach was to use the $\chi^2$ method, which minimizes the squared differences between synthetic and observed spectra using multiple pixels of the line core and wings. In this way, the $\chi^2$ method is less susceptible to noise or to light blends on the wings, and it can be regarded as the most stable option. In some cases, for instance, when the wings are heavily blended, it is more appropriate to use a method that ignores the wings, namely int (intensity) and wln (wavelength). When there was a clear line core, we used the int method, which averages the line depths of five points around the core and interpolates this against the synthetic spectra. Another possibility is to use the wln method, which takes the abundance that matches in the single pixel located at the exact wavelength of the line best. This was done, for example, when the line of interest was on the wing of a strong blending feature. \\
\indent Each method has a set of quality flags that account for issues in the derivation of the corresponding abundances. Our default choice was to only consider lines with good quality flags (flag = 1). In some cases, this criterion was weakened and lines were considered even though they were flagged. For instance, very strong lines are typically flagged by the eqw (equivalent width) method because there is an internal limit for the line depth. When the line is clean, no other flag is raised, and the result is consistent with the remaining four methods. We therefore used the line in the averaging for the final abundance. \\
\indent The final decision to use or reject a line was only taken after visual inspection. We selected the method and generally relied on the quality flags, but also focused on continuum mismatches. If necessary, continuum shifts were performed manually. \\
\indent We determined the abundance in each of the four targets by first adjusting the convolution parameter necessary for the match between observed and synthetic spectra. Because they are strong and clear, we used the Fe lines for this task. After the convolution parameter was fixed, we performed one iteration over Fe to obtain the metallicity and three iterations over C, O, and N to fit the molecular features. The Fe and molecular bands strongly affect the overall shape of the spectra. Additionally, we fixed the $\isotope[12]{C}/\isotope[13]{C}$ ratio to 5 because this was the lower limit we estimated by means of the $\isotope[13]{CH}$ and $\isotope[13]{CN}$ features near \unit[4048.3]{\angstrom} \citep{masseron14CH} and \unit[8004.7]{\angstrom} \citep{sneden14CN}, respectively. An accurate measurement could not be performed, but the estimated lower limit of 5 is consistent with the red giant evolutionary phase of the stars. The next step consisted of calculating the abundances of Ti, V, Cr, Mn, Co, and Ni because these elements have many lines in the optical and consequently often contaminate lines of interest. This was followed by the calculation of other light (Z $\leq30$) and heavy elements (Z $> 30$) with many lines, such as Si, Ce, and Nd, to further account for blends and to reproduce the atmosphere of the stars as well as possible before we determined the abundances of the species Rb-Pb, which are critical for constraining nucleosynthetic scenarios. When no abundance measurement was possible, we estimated an upper limit consistent with all lines by visually comparing the synthetic and measured spectra. \\
\indent The final abundance value of each species is then given by the average over the abundances from all accepted lines scaled\footnoteref{Bracket} to the solar zeropoints from \citet{Grevesse2007}. In Table \ref{table:abuheavy} we list the final abundances of the heavy (Z $> 30$) elements averaged over all, neutral, and ionized lines, the abundances corresponding to each ionization state, the line-by-line standard deviation, and the number of lines used. The discussion in Sect. \ref{models} is based on the ionization-independent average over all lines. \\
\indent Analogous results for the light (Z $\leq30$) elements are available at the CDS. A detailed comparison between the results of \citetalias{brauner23} and and the abundances of the light elements is beyond the scope of this work. However, we briefly mention some possible sources of discrepancies. The abundance differences in C, N, and O, can be explained by a  slight difference in the calculation procedure. That is, in \citetalias{brauner23}, the default approach to determine the C, N, and O abundances was to perform three iterations over these elements in order to achieve molecular equilibrium, but to rely on the automatic update of the abundances after each iteration step. This automatic update consists of averaging the abundances over all the lines with good quality flags, disregarding possible poor fits that can only be identified by a visual inspection. Furthermore, only the results from the $\chi^2$ method were used. This was done to facilitate an automated calculation of background stars for comparison. In the present work, the abundances were updated in each step only after such visual inspection, and as mentioned above, we did not limit ourselves to the $\chi^2$ method either. For the remaining elements, the differences in the abundances are most likely due to the line selection. More precisely, the number of considered lines in this work is in general higher, resulting in a more robust average line-to-line abundance. Moreover, the susceptibility to derivations from local thermodynamic equilibrium is not the same for lines in the infrared and the optical and may cause discrepancies in the derived abundances. This is especially true for Na and Al. Nevertheless, the claims about enhancements in the light elements O, Al, and Si, as well as in the overall pattern found by \citetalias{brauner23}, can be confirmed.

\subsection{Uncertainties}
The uncertainty associated with the final abundances has two parts: the random error, represented by the line-to-line scatter, and the systematic error, which reflects the uncertainty related to errors in the stellar parameters. The former is specified by the standard deviation $\sigma_{std}$ of the abundances derived from each line. When no standard deviation exists, that is, when the measurement is based on only one line, we fixed it to be $\sigma_{std}$=0.15. Furthermore, we defined a minimum standard deviation by imposing a lowest value of $\sigma_{std}$=0.10 in the case of two lines and $\sigma_{std}$=0.05 in the case of three or more lines. \\
\indent For the systematic error, we followed the procedure described in \citetalias{brauner23} and focused on the heavy (Z $> 30$) elements. We performed the calculation on 2M00044180 because this star has the fewest upper limits. For the uncertainty in the Ga I abundance, 2M22480199 was used. We determined the sensitivity of the abundances to typical deviations of $\pm$\unit[100]{K}, $\pm$\unit[0.3]{dex}, and $\pm$\unit[0.05]{km/s} in T$_{eff}$, $\log$ g, and microturbulence velocity $\mu_t$, respectively. Modifications on the line lists (available at the CDS) and continuum adjustments were maintained for the error calculation. The obtained deviations that are equivalent to the $\sigma_{param,[X/Fe]}$ from \citetalias{brauner23} are also available at the CDS. The total uncertainty was calculated by taking the square root of the sum of the squared $\sigma_{param,[X/Fe]}$. We note that, overall, a deviation in T$_{eff}$ affects the uncertainty most, followed by $\log$ g.

\section{Discussion\label{discussion}}
In this section, we discuss the potential binarity of the four stars and the possible origin(s) of the observed abundance patterns. 

\begin{table*}
\caption{Radial velocity data from \textit{Gaia} DR3 for the four targets.} 
\label{table:RVgaia}      
\centering                          
\begin{tabular}{c c c c c c c}        
\hline\hline
    Star & Combined RV & \# transits & Time coverage & PtoP RV amplitude Gaia & PtoP RV amplitude UVES & Variable? \\ 
    & [km/s] &  & [days] & [km/s] & [km/s] & \\
    \hline
    2M00044180 & -59.94$\pm$0.48 & 35 & 1020.90 & 8.71 & 2.95 & No \\
    2M13472354 & 32.75$\pm$0.89 & 19 & 906.52 & 12.74 & 1.39 & No \\
    2M18453994 & 176.96$\pm$0.74 & 9 & 778.57 & 5.83 & 1.98 & No \\
    2M22480199 & -211.68$\pm$1.17 & 11 & 899.59 & 11.62 & 8.64 & No \\ \hline
\end{tabular}
\tablefoot{Combined RV (median of the epoch RV time series), number of transits, time coverage of the time series, and peak-to-peak amplitude of the RV time series for \textit{Gaia} DR3 and UVES (V$_{\gamma}$) from Table \ref{table:obslog}. The last column shows the result from applying the variability criterion (see main text).}
\end{table*}

\subsection{Binarity\label{binarity}}
The possibility that the P-rich stars are part of binary systems has been discussed in previous studies (\citealp{masseron20a}, \citetalias{brauner23}). We revisit this consideration because we note that the radial velocity (RV) variations shown by the UVES observations (Table \ref{table:obslog}) differ from the scatter found by the APOGEE-2 survey. In the case of 2M00044180, the data from APOGEE-2 reveal a low RV scatter (\unit[0.079]{km/s}) based on ten observations collected with a maximum time lag of 2600 days. This large number of observations that are well separated in time was deemed a reliable exclusion criterion for binarity in \citetalias{brauner23}. However, the RV values in Table \ref{table:obslog} show an appreciable scatter during one month for this target. A similar scatter can be observed for 2M22480199, where the low RV scatter (\unit[0.194]{km/s}) from APOGEE-2 is not reliable because the RV was measured only three times on consecutive days. \\
\indent In contrast, the other two stars, 2M13472354 and 2M18453994, do not show significant RV scatter, but this observation cannot be confirmed by the APOGEE-2 results because 2M13472354 has a low RV scatter (\unit[0.160]{km/s}), but was observed three times within only six days and 2M18453994 was only observed once. We therefore have no information on its RV scatter. \\
\indent The \textit{Gaia} mission provides in its third data release (DR3) \citep{gaiaedr3} combined RV measurements from 34 months of observations, including the targets studied in this work. In Table \ref{table:RVgaia}, we summarize the combined RVs, number of transits, time coverage of the time series, and the peak-to-peak amplitude of the time series. The values of the combined RV agree well with the RVs V$_{\gamma}$ from the UVES observations presented in Table \ref{table:obslog} when the errors and peak-to-peak amplitudes are taken into account. Table \ref{table:RVgaia} also shows the result from applying a criterion for variability, which is based on two variability indices: the probability value for consistency (\texttt{rv\_chisq\_pvalue}), and the renormalized goodness of fit (\texttt{rv\_renormalised\_gof}) of the time series. \citet{katz23} proposed the criterion that stars with \texttt{rv\_chisq\_pvalue} $\leq$ 0.01 and a \texttt{rv\_renormalised\_gof} > 4 can be classified as variable. They recommend applying this criterion to targets with more than ten transits and a template used for the cross-correlation with T$_{eff}$ between \unit[3900]{K} and \unit[8000]{K}. Both recommendations are fulfilled for the observations of our targets, except for 2M18453994, which has only nine transits. When this criterion is applied, none of the four stars shows variability in RV. This conclusion is supported by the number of transits and large time coverage. However, the relatively high change in a short time of the peak-to-peak amplitude for 2M22480199 obtained from the UVES spectra (see Table \ref{table:obslog}) contradicts the results from \textit{Gaia} DR3 when the criterion from \citet{katz23} is applied. 
We therefore suggest more multi-epoch RV measurements using high-accuracy spectroscopic instruments to monitor the variations in our four targets and other P-rich stars in order to discard or confirm binarity, and following from this, a possible mass-transfer scenario.

\begin{table*}[!t]
\caption{Observed elements that depart by \unit[0.3]{dex} or more from the comparison with theoretical simulations.}
\label{tab: summary}      
\centering
\begin{tabular}{r | c | c | c | c | c | c }        
\hline\hline                 
 & \multicolumn{3}{c}{Scenario 1} & \multicolumn{2}{c}{Scenario 2} & \multicolumn{1}{c}{Scenario 3}  \\    
\hline
 & 1a & 1b & 1c & 2a & 2b & \\
 \hline\hline                 
 
2M22480199  & \textbf{Rb}, Ag, Ba, \textbf{La} & \textbf{first peak} & Ge, \textbf{Sn}, Ba & Ga, \textbf{Rb}, Ag, \textbf{La} & \cellcolor{green!25} Ga, \textbf{Rb}, Ag, Ba &  \cellcolor{green!25} \textbf{Sn} \\ 
 &  &  &  & & \cellcolor{green!25} & \cellcolor{green!25} \\
 \hline
2M18453994  &  \textbf{Rb}, Y, Nb, Ag, & \textbf{first peak} & \textbf{Sr}, Y, Mo, Ru, &  \textbf{Rb, Y}, Nb-Ag, & \textbf{Rb, Y}, Nb-Ag & \textbf{Sr, Y}, Rh-Cs, \\
 & Ba, \textbf{La-Eu}, Dy & \textbf{Pr, Sm}, Hf, \textbf{Pb} & Ba, \textbf{La-Eu}, Dy &  \textbf{Pr, Sm}, Hf, \textbf{Pb} & \textbf{Cs} & Hf \\
 \hline
2M13472354   & Ga, \textbf{Sr}, Mo, Ru,  & \textbf{first peak} & Ba, \textbf{La}, Hf & Ga, Nb, Mo-Ag & Ga, Mo, Ru, Ag, & \cellcolor{green!25} Pr, \textbf{Sm} \\
& Ba, \textbf{La}, Hf &  Pr &   & Pr & Ba, Pr, \textbf{Sm} & \cellcolor{green!25} \\
 \hline
2M00044180  & \textbf{Rb, Y}, Nb, Ag,  &  \textbf{Rb-Ru} & \cellcolor{green!25} Ag, Ba & \textbf{Rb}, Nb, Ag & \textbf{Rb}, Nb, Ag, Ba & \cellcolor{green!25} Ag, Ba, \textbf{Pb} \\
& Ba, \textbf{La} &  & \cellcolor{green!25} & & & \cellcolor{green!25} \\
\hline          
\end{tabular}
\tablefoot{Scenario 1a (single i-process exposure producing mostly the Zr region, using Zr as the reference for the dilution), 1b (single i-process exposure producing mostly the Ba region, using Ba as a reference element), 1c (single s-process exposure, using Zr as reference element), 2a (double i-process exposure, normalizing to Zr and Ba the two components), 2b (the same as 2a, but using La as reference instead of Ba), and scenario 3 (s-process signature and i-process component at the second peak starting from s-process seeds with La as reference). Elements in bold are classified as reliably measured. No scenario works perfectly for all the available abundances, but we highlight in green the solutions that show the least discrepancies.}
\end{table*}

\subsection{Comparison with the predictions from stellar nucleosynthesis models\label{models}}

 In previous works, the new class of P-rich stars was identified based on their peculiar elemental abundance pattern of intermediate-mass elements lighter than Fe. \cite{masseron20a} and \cite{masseron20b} highlighted anomalous enrichments for elements above Fe in 2M13535604+4437076 and 2M22045404-1148287. In particular, for the elements available in the Rb-Mo (Z=37$-$42) mass region, for the lanthanides (Z=58$-$71) and up to Pb (Z=82), 2M22045404-1148287 showed typical enhancements compared to Fe of up to an order of magnitude, with the relevant exception of an enrichment by two orders of magnitude measured for Ba \citep[][]{masseron20b}. The reported stellar abundances were not compatible with an r-process pattern, indicating that the r-process is not the main source of the observed anomalies. We can derive similar conclusions for the additional stars reported in this work (2M00044180-0005553, 2M13472354+2210562, 2M18453994-3010465, and 2M22480199+1411329) with more measured elements: Eu and other typical r-process elements are not particularly boosted compared to nearby elements that are not efficiently produced in r-process conditions. Therefore, we can safely exclude the scenario in which this contribution explains the abundance distributions of P-rich stars beyond Fe. \\
\indent \cite{masseron20b} identified some similarities with the s-process signature in AGB stars, for instance, a subsolar [Rb/Sr] ratio, which is associated with the typical low neutron density conditions in these stars \citep[e.g.,][]{abia:01}. On the other hand, the observed [Ba/La] ratio, which is much higher than solar, was not compatible with the s-process. As shown in Sect. \ref{Rb}, both aspects are also true for the new targets. Other arguments that would argue against the scenario in which AGB stars cause the abundances in P-rich stars are their relatively normal or low C abundances with respect to O (where C is expected to be also enhanced together with the s-process elements in AGB stars), together with the low RV scatter observed for a significant fraction of the P-rich stars, indicating that they are not part of close binary systems \citepalias[e.g.,][and Section \ref{binarity}]{brauner23}. \\
\indent The i-process has been claimed to be observed in low-metallicity stars and post-AGBs \citep[][]{bertolli:13,dardelet:14, mishenina:15, hampel:16, hampel19} and represents a promising alternative to explain the neutron-capture pattern of the P-rich stars. Simulations of the i-process are available for low-metallicity AGB stars \citep[e.g.,][]{choplin:21} and rapidly accreting white dwarfs \citep[][]{denissenkov17, denissenkov:19}, and they match the neutron-capture elemental pattern of a subclass of carbon-enhanced metal-poor stars (CEMP-i) relatively well. However, in addition to the clear discrepancy of the carbon abundance in the two classes of stars, \citet{masseron20b} showed that the observed neutron-capture abundance pattern of the P-rich stars is different from that of the CEMP-i, meaning that other i-process sites have to be considered. \cite{hampel19} considered the potential i-process signature of Magellanic post-AGB stars for metallicities higher than those of metal-poor CEMP-i stars. It is difficult to compare our data with the observed abundance patterns discussed by \cite{hampel19}, however. We show below that the most challenging constraints for stellar simulations indeed come from the elements observed between Zr (Z=40) and Ba (Z=56), which are not available for the sample of post-AGB stars to compare with nucleosynthesis calculations. \\
\indent In this section, we compare the abundance pattern of heavy elements measured for 2M00044180, 2M13472354, 2M18453994, and 2M22480199 with i-process nucleosynthesis calculations and with s-process simulations for fast-rotating massive stars. For the i-process, we used the same setup as adopted for previous comparisons with metal-poor stars \citep[e.g.,][]{bertolli:13, roederer:16, roederer:22, ji:24}. Based on the considerations made in this section and more in general by \citet{masseron20a} and \citetalias{brauner23}, we would rather consider massive stars as the most likely candidates to generate the anomalous abundance pattern in P-rich stars. However, while it has been proposed that massive stars might host the i-process \citep[][]{clarkson:18, banerjee:18, clarkson:21}, the main properties of these types of events in massive stars are currently extremely uncertain, and we lack comprehensive multidimensional hydrodynamic models, which are necessary to properly constrain the shaping of the He-burning material in which the i-process will eventually take place \citep[e.g.,][]{herwig:14, woodward:15}. For these reasons, we adopted a simplified single-zone trajectory in this computational experiment, which is intended to qualitatively explore the range of possible i-process elemental ratios \citep[see e.g.,][]{roederer:22, ji:24}. \\
\indent For the s-process in metal-poor fast-rotating massive stars, we used the same setup as adopted previously by, for example, \cite{hirschi:08, best:13, roederer:22}. The post-processing network code PPN was used to generate the nucleosynthesis calculations \citep[e.g.,][]{pignatari:12}. The single-zone trajectory was extracted from a complete \unit[25]{M$_\odot$} star \citep[e.g.,][]{hirschi:08}. In order to take the primary $^{22}$Ne production arising from stellar rotation into account, the initial abundance of $^{22}$Ne was set to be 1\%, in agreement with \cite{pignatari:13}, for example. We considered as initial metallicity for the simulations [Fe/H] = \unit[-1.5]{dex}, which is consistent with the range of metallicity of the P-rich stars. \\
\indent More specifically, three different nucleosynthesis scenarios were considered: a single neutron exposure due to either the i-process (1a, 1b) or the s-process (1c), and a composition of two i-process neutron exposures that separately feed the Zr region and the Ba region (2a, 2b). We also tested the case of an s-process neutron-exposure feeding the first peak and a second i-process event producing heavier elements starting from s-process seeds (3). The results of these computational experiments are summarized in Table \ref{tab: summary}, where we provide a list of the elements (or element mass regions) that depart in observations and predictions by about \unit[0.3]{dex} or larger (i.e., outside the observed error bars). As we mentioned earlier, we no longer considered the r-process scenario and the s-process in AGB stars because these two options were discarded as a solution in previous works (\citealp{masseron20a,masseron20b} and \citetalias{brauner23}). The selected results are presented in Figs. \ref{fig: scenario1_2M22480199+1411329}-\ref{fig: scenario3_all}.

\subsubsection{Scenario 1: Single i-process or s-process exposure with initial abundance variations \label{subsec: iprocess_1}}

\begin{figure*}[hbt!]
  \includegraphics[width=0.5\textwidth]{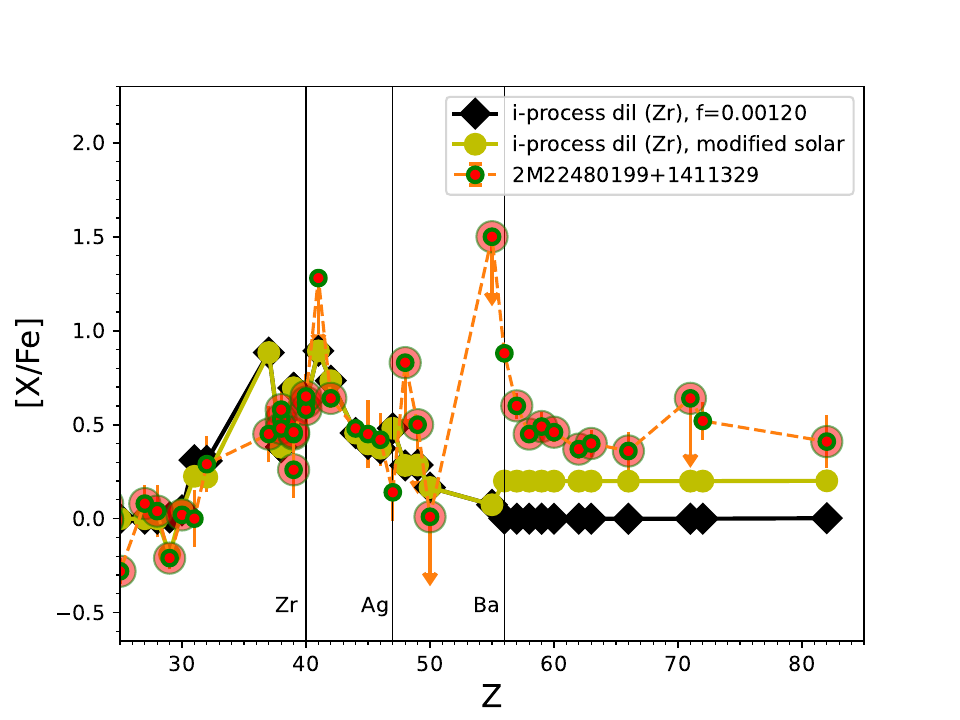}
  \includegraphics[width=0.5\textwidth]{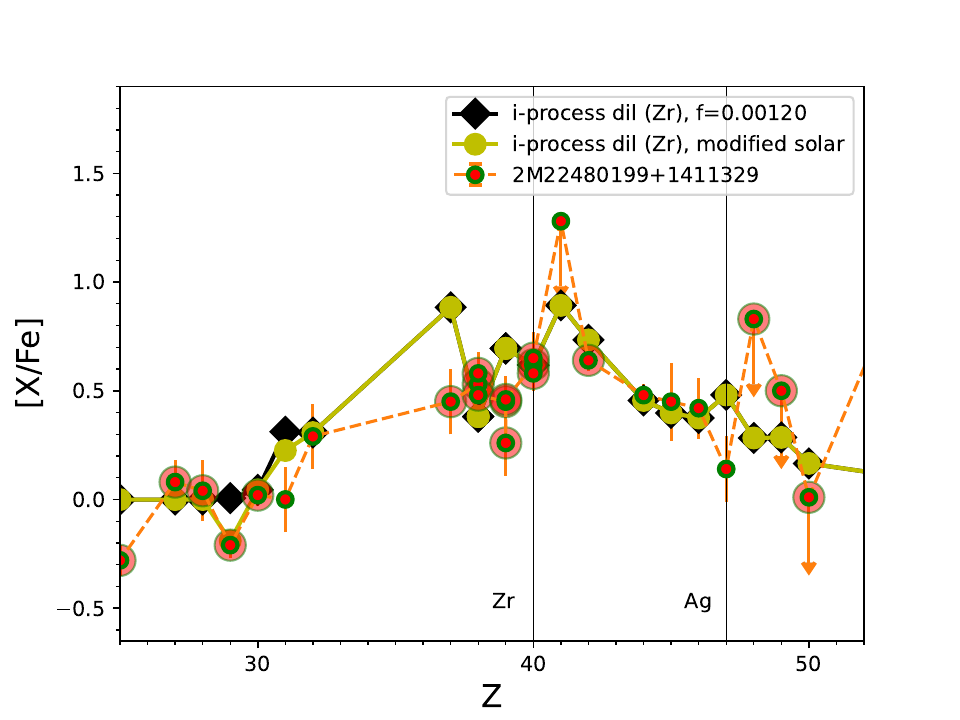}
  \includegraphics[width=0.5\textwidth]{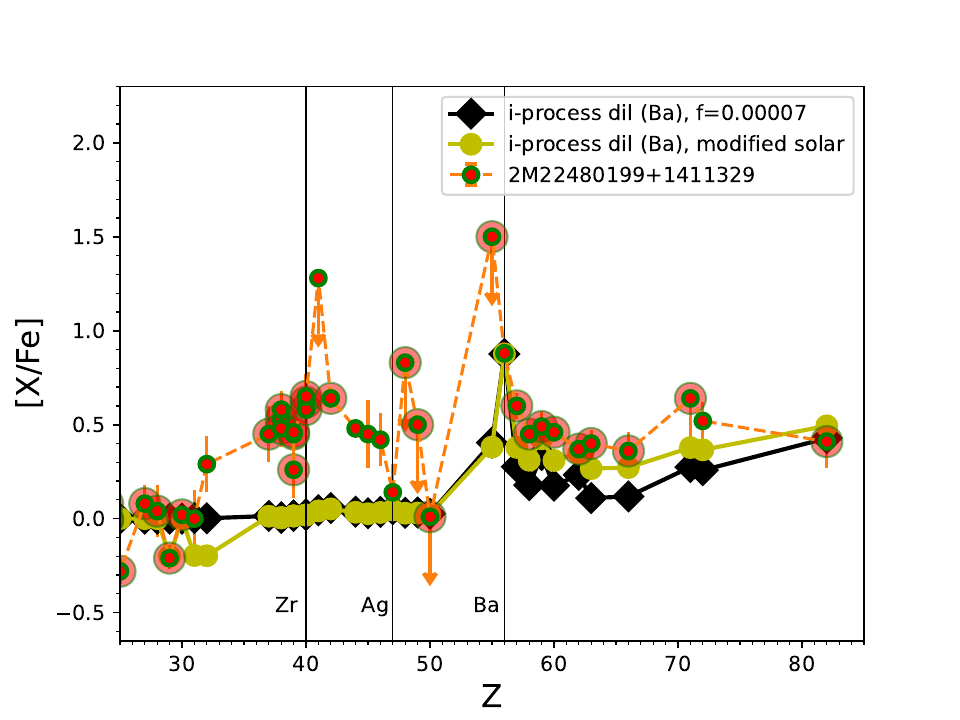}
  \includegraphics[width=0.5\textwidth]{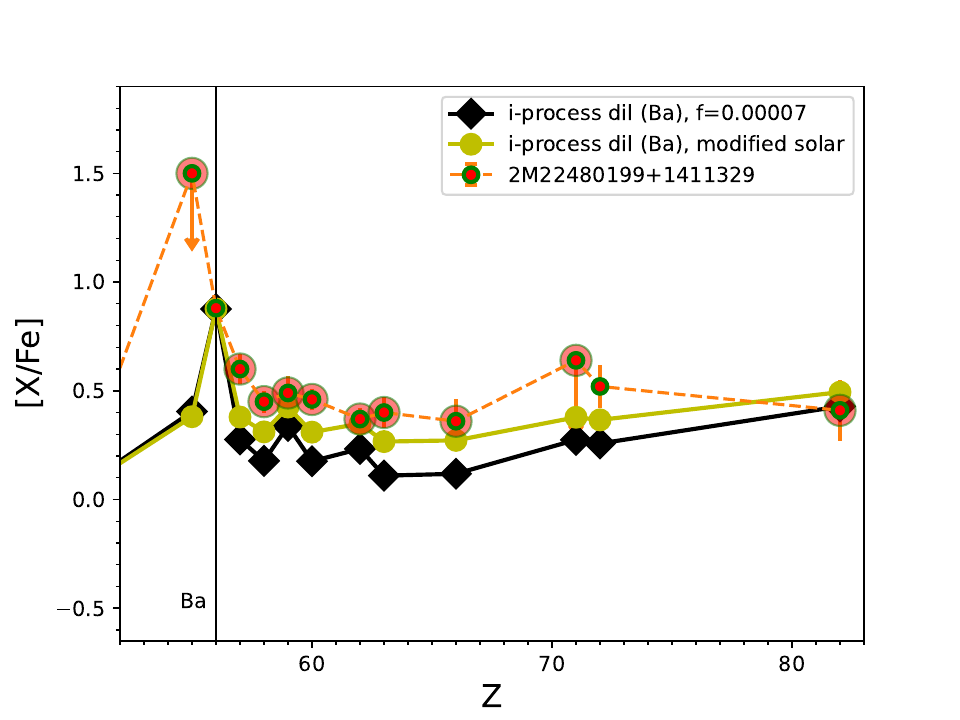}
  \includegraphics[width=0.5\textwidth]{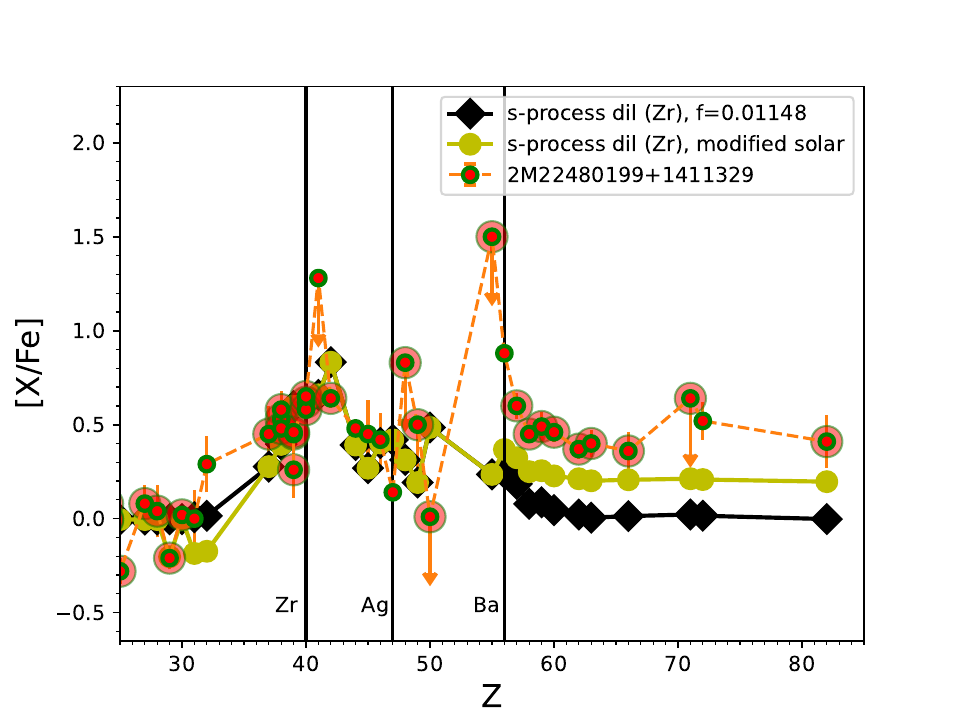}
  \includegraphics[width=0.5\textwidth]{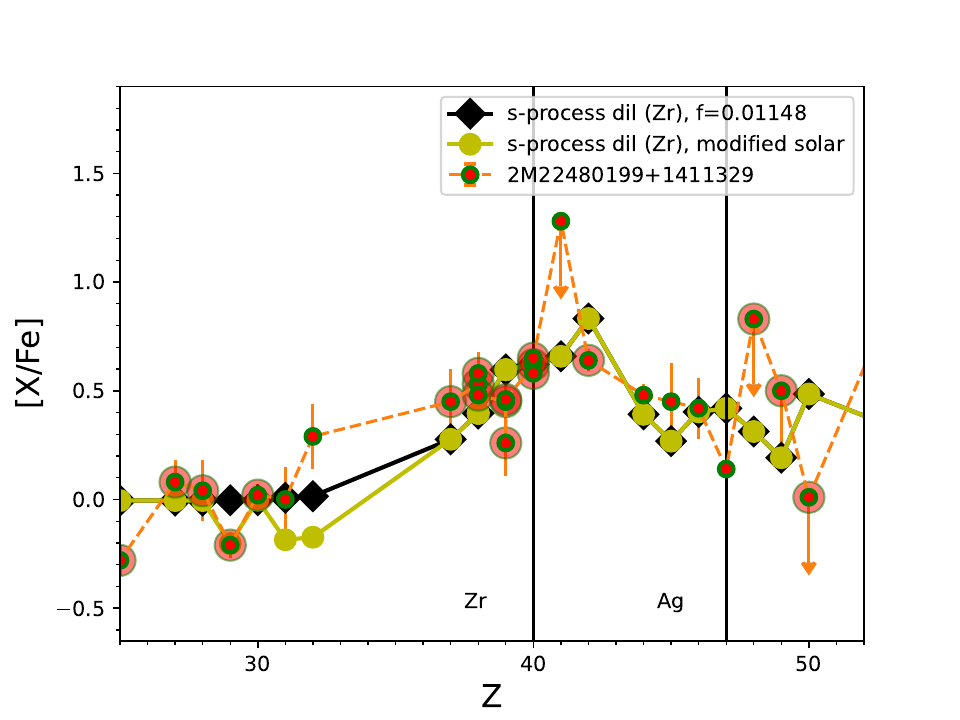}
  \caption{Complete elemental abundances beyond Fe of 2M22480199 (left panels) and zoomed over a smaller mass range (right panels) compared with i-process simulations that peak in the Zr region (1a, upper panels) and at Ba (1b, middle panels), and with s-process simulations from fast-rotating massive stars (1c, lower panels). The elements with the most reliable observational results (measurements or upper limits) are highlighted with large orange circles. Stellar simulations are diluted with pristine material to reproduce the abundance of the reference element Zr (upper and lower panels) or Ba (middle panels). The results from the dilution with two different pristine components are shown: solar scaled, and modified by \unit[0.2]{dex} from solar (see the text for details).   
  }
  \label{fig: scenario1_2M22480199+1411329}
\end{figure*}

For this first set of tests, we explored three nucleosynthesis subcases within the main scenario of a single event whose signature dominates the observations. For cases 1a and 1c, the observed abundances were compared with the yields of a single neutron exposure that mostly feed the light neutron-magic peak at Zr (Z=40) and are due to either the i- or the s-process, respectively. In case 1b, we considered a single i-process exposure that feeds the second neutron-magic peak at Ba (Z=56). The simulation results were mixed with pristine gas, where the dilution factor f was given by the fraction of pure nucleosynthesis yields diluted with (1 - f) of normal material, which is required to reproduce the observed abundance of Zr (case 1a and 1c) or Ba (case 1b). \\
\indent We also explored the impact on the results of diluting with a non-solar pristine gas, that is, with enhancements of Ba and heavier elements by \unit[0.2]{dex}, and a depletion of Cu (Z=29), Ga (Z=31), and Ge (Z=32) by \unit[0.2]{dex}. These variations would be consistent with the abundance scatter of nonevolved stars of the Milky Way disk with the same metallicity of the stars discussed here. \\
\indent Case 1a and case 1b may represent a scenario in which single massive stars would be the source of the i-process signature, following a single H ingestion event in the He-burning shell \citep[e.g.,][]{roederer:16,ji:24}. Case 1c, on the other hand, tests the assumption that the s-process in fast-rotating massive stars would generate the abundances \citep[e.g.,][]{pignatari:08, barbuy:21}.

\paragraph{2M22480199+1411329} For case 1a (Fig. \ref{fig: scenario1_2M22480199+1411329}, upper panels), 2M22480199 matches well up to Sn (Z=50) overall, but overproduces Rb (Z=37) and Ag (Z=47) by about \unit[0.4]{dex}. The Rb overproduction could be mitigated by using a larger neutron exposure, but this would strongly deteriorate the fitting beyond Mo (Z=42) up to the Sn upper limit. The abundances for Ba and beyond are not reproduced. Instead, the anomalies beyond La are roughly flat, which could be explained by an enriched pristine composition of heavy elements beyond Ba with respect to Fe. In the figure, we show the impact of an enrichment of about \unit[0.2]{dex} (modified solar), which would be almost compatible with observations within \unit[0.3]{dex}, except for an overproduction of La (by \unit[0.4]{dex}) and Ba (by \unit[0.7]{dex}). We showed in Sect. \ref{Ba} that the lack of non-LTE (or NLTE) corrections for Ba introduce some uncertainties that might affect the current abundance evaluation. For this reason, and due to the extreme overabundance, Ba is preliminarily listed as not reliable in Table \ref{tab: summary}. Overall, La is probably more reliable. We currently consider it unlikely that observations beyond Ba in our stars are entirely driven by anomalies in the interstellar medium for all the P-rich stars. For instance, this scenario would not be feasible for the other P-rich stars discussed in the past, 2M13535604 and 2M22045404, where much larger anomalies are observed. For the specific case of 2M22480199, an independent confirmation of the Ba abundances would be helpful to confidently exclude case 1a. Similar considerations apply for case 1c (Fig. \ref{fig: scenario1_2M22480199+1411329}, lower panels). An overproduction of the Sn upper limit by about \unit[0.5]{dex} is the only significant discrepancy observed compared to the s-process distribution (when we exclude Ba, as discussed before). Several lighter elements would be overproduced for a lower neutron exposure to reproduce Sn. Minor differences of about \unit[0.3]{dex} are found for Ag and Hf (Z=72). In case 1b (Fig. \ref{fig: scenario1_2M22480199+1411329}, central panels), the upper limits mentioned above and of the elements beyond Ba match well. However, no significant contribution to lighter elements between Rb and Pd (Z=46) is shown.

\paragraph{2M18453994-3010465} This star is not consistent with the i-process simulations in either case 1a or case 1b (Fig. \ref{fig: scenario1_2M18453994-3010465}, upper panels and middle panels). For case 1a, the low upper limits for Rh (Z=45), Pd (Z=46), and Ag (Z=47) compared to the production of Zr and Mo in particular would be consistent with a strong Rb production, which is not observed. In particular, Rb, Y (Z=39), and Nb (Z=41) are overestimated by \unit[0.6]{dex}, \unit[0.5]{dex}, and \unit[0.4]{dex}, respectively, and the Ag upper limit is overestimated by about \unit[0.6]{dex}. In case 1b, the low production below Ba satisfies the upper limits mentioned above, but Pr (Z=59), Sm (Z=62), Hf (Z=72), and Pb (Z=82) are overproduced. In this case, the elements at the first neutron-magic peak are not produced significantly, either. As we showed for case 1a and similarly for case 1c (Fig. \ref{fig: scenario1_2M18453994-3010465}, lower panels), the upper limits of Rh, Pd, and Ag force us to use a neutron exposure that leads to an overproduction of lighter elements: Sr and Y are overproduced by \unit[0.5]{dex} and \unit[0.6]{dex}, respectively, and Mo and Ru (Z=44) are underproduced by \unit[0.3]{dex} and \unit[0.4]{dex}, respectively. Furthermore, the whole region from Ba to Dy (Z=56$-$66) is not reproduced. The production of Ba is missed by \unit[1.6]{dex} and the region from La to Dy region by \unit[0.3--0.4]{dex} in cases 1a and 1c.

\paragraph{2M13472354+2210562} For case 1a (Fig. \ref{fig: scenario1_2M13472354+2210562}, upper panels), the i-process nucleosynthesis distribution up to Ba overestimates Ga (Z=31), Mo (Z=42), and Ru (Z=44) by \unit[0.3]{dex} and underproduces Sr by \unit[0.3]{dex}. Ba is strongly underestimated, and La and Hf (Z=72) are underestimated by \unit[0.35]{dex} and \unit[0.3]{dex}, respectively. In case 1b (Fig. \ref{fig: scenario1_2M13472354+2210562}, central panels), the i-process pattern underestimates Sr, Y, and Zr by \unit[0.7--0.5]{dex}, but fits the remaining species between Mo and Pb (Z=82) quite well, except for Pr (Z=59). In case 1c (Fig. \ref{fig: scenario1_2M13472354+2210562}, lower panels), the s-process pattern is remarkably consistent with observations up to the Sn (Z=50) upper limit. However, Ba and La are still underestimated by \unit[1.4]{dex} and \unit[0.4]{dex}.

\paragraph{2M00044180-0005553} Case 1a (Fig. \ref{fig: scenario1_2M00044180-0005553}, upper panels) shows an overproduction of Rb (Z=37), Y (Z=39), and Nb (Z=41) of \unit[0.4--0.3]{dex}, and Ag is overproduced by \unit[0.6]{dex}. As we discussed earlier, Ag and the elements in this region yield powerful constraints through a comparison of stellar nucleosynthesis simulations with observations, where the abundances are lower than in the Zr and Ba peak. Case 1b (Fig. \ref{fig: scenario1_2M00044180-0005553}, middle panels) underproduces the Rb-Ru (Z=37$-$44) region by up to \unit[0.7]{dex}, while the rest of the distribution compares well. Case 1c shows that the s-process in fast-rotating massive stars could explain the observed distribution up to the Sn (Z=50) upper limit, except for the subsolar [Ag/Fe] ratio, which is overproduced by \unit[0.5]{dex}. The heavier elements are also compatible with the s-process predictions, except for Ba (underproduced by \unit[0.8]{dex}).

\subsubsection{Scenario 2: Double i-process exposure with initial abundance variations \label{subsec: iprocess_2}}

The second scenario is based on the assumption that two distinct i-process events have generated the observed abundance patterns, with different neutron exposures and with distinct dilution factors. \cite{koch:19} reported the signature of a double i-process neutron exposure in the anomalous abundances of the metal-poor CH-star 10464. In this context, it might be argued that the double exposure is due to two following H-ingestion events or to two different outcomes in different parts of the He shell initiated from the same H ingestion event \citep[e.g.,][]{clarkson:21}. We cannot exclude either of these two options because of the current stellar uncertainties. To select the neutron exposures that provide the best fit to the observations, we normalized the first component to Zr and the second component to Ba (case 2a) or to La (case 2b). In case 2b, we did not consider the Ba abundance for the fit because of the uncertainty caused by lacking NLTE corrections for Ba.

\paragraph{2M22480199+1411329} In case 2a (Fig. \ref{fig: scenario2_2M22480199+1411329}, left panel), the dual i-process component shows a mild overproduction of Ga (Z=31, \unit[0.3]{dex}) and an overproduction of Rb (Z=37, \unit[0.5]{dex}) and Ag (about \unit[0.35]{dex}). As in the previous section, this demonstrates that our measurements in the Rh-Ag (Z=45$-$47) region provide a valuable constraint to challenge nucleosynthesis calculations. Within the variations considered for the composition of the pristine gas, the predicted La is underproduced by about \unit[0.35]{dex}, and several elements up to Hf (Z=72), are underproduced by \unit[0.2--0.3]{dex}. For elements beyond Ba, measurements in the Ag-Sn (Z=47$-$50) region and Pb (Z=82) constrain the allowed range of i-process neutron exposures. Case 2b (Fig. \ref{fig: scenario2_2M22480199+1411329}, right panel) also shows problematic Rb, Ag, and Ba (all with \unit[0.5]{dex} overproduction), while the fit is good for the other elements. Ag is slightly worse compared to case 2a because some amount of this element is also produced by the i-process component with the largest neutron exposure here. 

\paragraph{2M18453994-3010465} This scenario looks unlikely for this star. For case 2a (Fig. \ref{fig: scenario2_2M18453994-3010465}, left panel), Rb, Y, and the whole Nb-Ag (Z=41$-$47) region are overproduced compared to observations by at least \unit[0.5]{dex}. For the region beyond Ba, the Cs (Z=55) upper limit combined with the high Ba/La ratio forces an overproduction of Hf (Z=72) and Pb (Z=82) by about \unit[0.5]{dex}. For case 2b (Fig. \ref{fig: scenario2_2M18453994-3010465}, right panel), similar issues are found up to Ag. Using La as reference element, we obtain a good pattern between Ba and Pb, but Cs is overproduced by an order of magnitude compared to the observed upper limit. 

\paragraph{2M13472354+2210562} For case 2a (Fig. \ref{fig: scenario2_2M13472354+2210562}, left panel), Ga (Z=31) and Nb (Z=41) are overproduced by \unit[0.4]{dex} and \unit[0.3]{dex}, respectively, as well as all the elements in the region between Mo and Ag (Z=42$-$47) by about \unit[0.5--0.7]{dex}. This is partially due to the combined contribution from the first and second i-process components. The second i-process component is overall consistent with 2M13472354, where the largest discrepancy is the overproduction of Pr (Z=59) by \unit[0.4]{dex}. Case 2b (Fig. \ref{fig: scenario2_2M13472354+2210562}, right panel) instead shows that when Ba is not considered, a slightly better match is found, but several problems remain: Ga, Mo, Ru (Z=44), and Ag are overproduced, Ba is underproduced by about \unit[0.4]{dex}, and Sm (Z=62) is overproduced by slightly less than \unit[0.4]{dex}.

\paragraph{2M00044180-0005553} Case 2a (Fig. \ref{fig: scenario2_2M00044180-0005553}, left panel) overproduces Rb (Z=37) and Nb (Z=41) by \unit[0.4]{dex} and Ag by \unit[0.7]{dex}. As we also showed in Sect. \ref{subsec: iprocess_1}, the subsolar Ag abundance measured in this star is challenging to reproduce by nucleosynthesis calculations. Beyond Ag, the obtained distribution matches the stellar pattern quite well. For case 2b (Fig. \ref{fig: scenario2_2M00044180-0005553}, right panel), we obtain similar results, except for Ba, which is now overproduced by \unit[0.35]{dex}.

\subsubsection{Scenario 3: Combination of s- and i-process exposures \label{subsec: iprocess_3}}

The i-process nucleosynthesis is activated by the $^{13}$C($\alpha$,n)$^{16}$O reaction in He-burning conditions, where the s-process may also take place in massive stars. Therefore, it is realistic to consider that the s- and i-process are both activated in the same stellar layers. For the range of metallicities considered here, the classical weak s-process is expected to be only marginally activated \citep[e.g.,][]{raiteri:92, baraffe:92}. Instead, in fast-rotating massive stars, the primary production of the neutron source $^{22}$Ne may boost the s-process production in the He core and He shell \citep[e.g.,][]{frischknecht:16}, which are the same regions as we considered for case 1c in Sect. \ref{subsec: iprocess_1}. In these conditions, the i-process could take place in the He shell where some s-process has been previously activated, and the s-process products could act as seeds of the i-process. Although this scenario is quite realistic to consider, given the common stellar environments in which the s-process and the i-process take place, our calculations here were aimed at deriving some constraints from the abundance behaviors defined by the nuclear physics properties. Arguments against this scenario are that, first, the s-process-rich ejecta could be dominated by the most internal C-shell material, which is independent of the s-process production in the He shell \citep[][]{frischknecht:16, limongi:18}. Second, the s-process production could be activated again after the i-process event and might affect the i-process abundance signature if the star has not yet reached the final collapse stage \citep[e.g.,][]{banerjee:18}. This is not captured by the simplified approach adopted here. Nevertheless, the effective capability of 1D stellar models to reproduce the stellar behavior during H ingestion events, and the following evolution of the He-burning stellar layers is also extremely uncertain. A careful benchmark with multidimensional hydrodynamic simulations is required. No core-collapse supernovae (CCSNe) models currently take these events properly into account, and therefore, there are no integrated stellar yields that could be used for a more consistent comparison. We also note that in the following tests, we derive the dilution factor for the final abundances from both the s-process component and the combined s-process and i-process component, but we do not show the abundance pattern with the total composition as in scenario 2. Since they most likely come from different parts of the CCSNe ejecta, unlike for the second scenario discussed in the previous section, it is unclear how we should add them in this case. This means that the two components have to be compared individually with the complete observed pattern.

\paragraph{2M22480199+1411329} In Fig. \ref{fig: scenario3_all} (top left panel), the s-process component underproduces Ge (Z=32) by \unit[0.3]{dex} and overproduces Ag and Sn (Z=50) by \unit[0.3]{dex} and \unit[0.5]{dex}, respectively. The i-process made by starting from s-process seeds reaches an efficient production of Ba and beyond with lower neutron exposures than the analogous cases in Sects. \ref{subsec: iprocess_1} and \ref{subsec: iprocess_2}, but because of this, it shows relatively more significant yields in the Ag-Sn (Z=47$-$50) region, which might be problematic here. For instance, in the case of 2M22480199, we overproduce the Sn upper limit by \unit[0.3]{dex}. We used La as a reference in this set of tests, and we therefore also overproduced Ba by \unit[0.3]{dex}. As we mentioned before, even upper limits in the mass region Pd-Sn (Z=46$-$50) appear to be extremely useful to constrain the nucleosynthesis scenario.

\paragraph{2M18453994-3010465} No reasonable pattern is found for this star with this scenario (Fig. \ref{fig: scenario3_all}, top right panel): The upper limits in the Pd-Ag (Z=46$-$47) region combined with the Zr abundances are compatible with low s-process neutron exposures, where Sr (Z=38) and Y (Z=39) are overproduced by about \unit[0.5]{dex}. At the same time, the high Ba and the low Pb (Z=82) abundances force the model to be above the Ag and Cs (Z=55) upper limits by about \unit[1.0]{dex}, among other difficulties.

\paragraph{2M13472354+2210562} As shown in Fig. \ref{fig: scenario3_all} (bottom left panel), the s-process pattern is consistent with the 2M13472354 abundances up to Sn (Z=50). Instead, the i-process seems to be overall consistent with the observed abundances beyond Ba, but overproduces Pr (Z=59) and Sm (Z=62) by \unit[0.5]{dex} and \unit[0.35]{dex}, respectively. 

\paragraph{2M00044180-0005553} Here, we obtain a good fit with the s-process up to Sn (Z=50), except for Ag, which is overproduced by about \unit[0.5]{dex} (Fig. \ref{fig: scenario3_all}, bottom right panel). As we discussed above, it is unlikely that the s-process achieves a high Zr production without already producing some Ag. The i-process overproduces Ag, Ba, and Pb (Z=82) by \unit[0.4]{dex}, \unit[0.3]{dex}, and \unit[0.4]{dex}, respectively. We cannot reduce these abundances by reducing the neutron exposure, because in this case, we would overproduce Cs (Z=55) and obtain an even higher Ag abundance.

\section{Summary and conclusions\label{conclusions}}

We have performed a detailed 1D LTE chemical abundance inventory of four P-rich stars based on high-resolution UVES spectra. The broad wavelength coverage from \unit[3300]{\angstrom} to \unit[9400]{\angstrom} allowed us to derive the abundances of 21 light and 27 heavy elements, including 59 individual species, complemented with upper limit estimates. The abundances of the light (Z $\leq$ 30) elements were used to define the atmospheric composition as accurately as possible and to account for possible blends in the lines of heavy elements (Z > 30), which are the focus of this work. Some of these elements, such as the elements from Rb to Sn, are rarely measured in stellar atmospheres. Our goal was to find observational clues for the nucleosynthetic process that causes the peculiar abundances of P-rich stars. Heavy-element abundance patterns provide valuable information for this goal. \\
\indent We found overabundances with respect to solar in the s-process-peak elements, accompanied by an extremely high Ba abundance and slight enhancements in some elements between Rb and Sn. Some of the upper limits that were estimated when a measurement was not possible are also constraining, such as in the cases of Cd I, In I, and Sn I. This demonstrates that constraining upper limits are very useful when evaluating different nucleosynthetic scenarios. We therefore encourage future efforts on providing upper limits in this elemental region. We classify our abundance calculation method as reliable because we fully leveraged the spectral range observed by using the maximum number of lines possible, verified each line fit by eye, and attended to the correct adjustment of blends. Overall, the abundances found here are compatible with previous results from \citet{masseron20b}. \\
\indent We emphasize the need for advances in NLTE studies covering the stellar parameters and elemental space suitable for the stars considered in this work, or P-rich stars in general. For instance, the abundance of Ba is of major interest to provide highly reliable constraints for nucleosynthetic models. Because we lack NLTE corrections, it has an associated uncertainty, although we attempted to keep it as small as possible by fitting the wings of the observed Ba lines. Many studies provide corrections for a limited number of elements and lines \citep[e.g.,][]{lind12,korotin15}, but with varying coverage of the stellar parameter range. An extended grid applied to the GALAH survey, including 13 elements in the optical, was recently presented by \citet{amarsi20}. No complete grid of NLTE corrections is available to date, however, that fulfills the requirements for a consistent and homogeneous analysis of giant stars that are diverse in terms of stellar parameters. This means that we can currently only provide a homogeneous analysis for 1D LTE. Another option to identify overabundances in the heavy elements is a differential approach with a consistent analysis in 1D LTE, which might alleviate the influence of 3D and NLTE effects. The drawback of a differential approach lies in the need for a large number of compatible comparison stars, that is, stars with similar stellar parameters and available spectra that ideally cover the same wavelength range. Nevertheless, for a comparison with theoretical models, absolute abundances and therefore NLTE correction are required. \\
\indent We have compared the observations with different sets of nucleosynthesis calculations for the s-process in fast-rotating massive stars and for the i-process. For 2M22480199, 2M13472354, and 2M00044180, the closest solutions seem to be given by a combination of the s-process that dominates the production of the elements below Ba and the i-process that produces most of the heavier elements (scenario 3). Alternative solutions were found for 2M22480199 from a combination of two i-process components (scenario 2b), and for 2M00044180 with a single s-process (scenario 1c). No solution was found for 2M18453994 because no scenario seems to match well. \\
\indent Galactic chemical evolution models that only include CCSNe for the production of P so far systematically underestimated the [P/Fe]-[Fe/H] trend (\citealp[see e.g.,][]{caffau11,cescutti12}), which indicates that other sources are not sufficiently taken into account. Very recently, \citet{bekki24} presented models of oxygen-neon (ONe) novae that successfully explain the evolution of [P/Fe] in our Galaxy. \citet{bekki24} also stated that in combination with certain conditions, including only a small amount of mixing with the ISM and the absence of strong pollution with CCSNe and SNe Ia products, newly formed stars could show a high [P/Fe] produced by ONe novae. When we consider this new and very recent scenario as a viable explanation for the origin of P-rich stars, there are two open questions to be addressed in the future: (i) whether the heavy-element pattern from the current study could also be reproduced by nucleosynthesis in these ONe novae models, and (ii) whether the subsequent formation of P-rich stars in the -2.0 < [Fe/H] < -1.0 range is possible given the timescale of the ONe novae P production. \\
\indent Large spectroscopic surveys such as APOGEE-2 are useful for a first estimation of the abundances of light elements and the mining of P-rich stars \citepalias{brauner23}. Complementary to this, we strongly encourage more high-resolution and high-quality observations in the optical range to access the abundances of heavy elements, as we did here. Furthermore, a larger group of P-rich stars with high-quality observations in the optical will help us to narrow down a statistically reliable fingerprint of these chemically peculiar stars in the heavy-element regime. We also recommend multi-epoch high-accuracy RV measurements (e.g., with ESPRESSO\footnote{Echelle SPectrograph for Rocky Exoplanets and Stable Spectroscopic Observations} or FIES\footnote{FIbre-fed Echelle Spectrograph}) to further investigate the binarity of the P-rich stars because their binarity can currently not be definitively ruled out.


\begin{acknowledgements}
    We thank the referee, Elisabetta Caffau, for the comments and suggestions that helped to improve this paper. 
    MB wishes to thank Nicola Caon for the IRAF support during the preparation phase.
    MB acknowledges financial  support  from  the European Union and the State Agency of Investigation of the Spanish Ministry of  Science  and  Innovation  (MICINN)  under  the  grant PRE-2020-095531 of the Severo Ochoa Program for the Training of Pre-Doc Researchers (FPI-SO). This project has been supported by the Lend\"ulet Program LP2023-10 of
    the Hungarian Academy of Sciences and by the NKFIH excellence grant TKP2021-NKTA-64.MP acknowledges the support to NuGrid from JINA-CEE (NSF Grant PHY-1430152) and STFC (through the University of Hull’s Consolidated Grant ST/R000840/1), and ongoing access to {\tt viper}, the University of Hull High Performance Computing Facility. MP and ML thank the European Union’s Horizon 2020 research and innovation programme (ChETEC-INFRA -- Project no. 101008324), and the IReNA network supported by US NSF AccelNet (Grant No. OISE-1927130).
    This work is based on observations collected at the European Southern Observatory under ESO programme 107.22RF.001. We thank the ESO astronomer for collecting the data. \\
    We made use of the VALD database, operated at Uppsala University, the Institute of Astronomy RAS in Moscow, and the University of Vienna. \\
    This research made use of computing time available on the high-performance computing systems at the Instituto de Astrofísica de Canarias. \\
    Software: The data processing of this work relies heavily on \verb|IRAF| \citep{tody86IRAF} and the Python libraries \verb|Pandas| \citep{mckinney2010pandas,reback2020pandas}, \verb|NumPy| \citep{harris2020array}, \verb|Astropy| \citep{astropy:2013,astropy:2018,astropy:2022} and \verb|Matplotlib| \citep{hunter07}.
\end{acknowledgements}

\bibliographystyle{aa} 
\bibliography{Refs} 

\clearpage

\begin{appendix}

\twocolumn

\section{Comments on elements\label{Comments}}

\begin{table}
\caption{Relevant abundance ratios.} 
\label{table:ratios}      
\centering                          
\begin{tabular}{c c c c}        
\hline\hline
    Star & $\text{[Rb/Sr]}$ & $\text{[Rb/Zr]}$ & $\text{[Ba/La]}$ \\ \hline
    2M00044180 & -0.13 & -0.14 & 0.54 \\
    2M13472354 & \dots & \dots & 1.05 \\
    2M18453994 & -0.13 & -0.15 & 1.02 \\
    2M22480199 & -0.05 & -0.17 & 0.28 \\ \hline
    2M13535604 & -0.38 & -0.31 & 0.76 \\ 
    2M22045404  & -0.08 & +0.04 &  0.76  \\
\hline
\end{tabular}
\tablefoot{For 2M13535604+4437076 and 2M22045404-1148287 the ratios are obtained with the data from \cite{masseron20b}.}
\end{table}

\begin{figure*}
  \resizebox{\hsize}{!}{\includegraphics{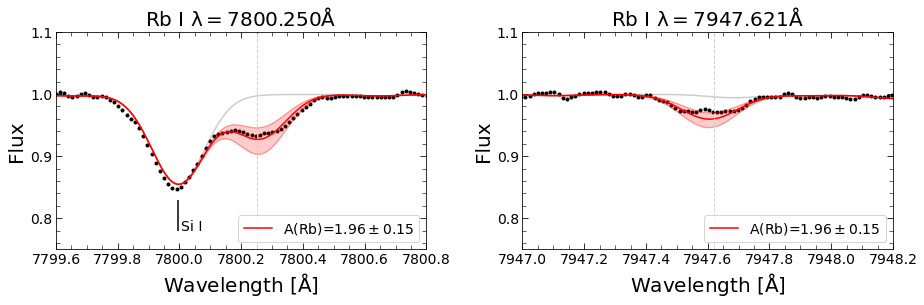}}
  \caption{Observed (black dots) and synthetic spectra of 2M00044180 around the two Rb I lines at \unit[7800.250]{\angstrom} (left panel) and \unit[7947.612]{\angstrom} (right panel). The synthetic spectra were computed for A(Rb)=$-\infty$ (gray line) and for the final abundance from Table \ref{table:abuheavy} (red line), given by the average abundance over the lines. The red shaded area corresponds to abundance variations within the amount of the standard deviation.}
  \label{fig:2M00Rb}
\end{figure*}

In the following, we discuss the abundance determination and results obtained for the heavy (Z $> 30$) elements presented in Table \ref{table:abuheavy}. 
We classify the abundances of light elements (Z $\leq30$) as generally reliable because of the large number of lines available in the optical. Some of these elements, for example Cu and Zn \citep[see][]{masseron20b}, may carry information on the P-rich star progenitor, but they require a separate analysis which is beyond the scope of this work. Therefore, we do not comment further on the abundance determination of the light elements, except for Fe. Given the discrepancy in the [Fe/H] value for 2M18453994 (see Table \ref{table:targetinfo}), we take a closer look at the determination of the Fe abundance which is used as metallicity proxy. In this context, we mention the possible influence of deviations from local thermodynamic equilibrium on the abundance results of Fe and other elements, but we recall that we opt for a standard 1D LTE approach, given that there are no grids of NLTE corrections for all elements and stellar parameters studied here. \\
\indent For the graphic discussion in the next subsections (Figs. \ref{fig:2M00Rb}-\ref{fig:2M00Pb}) we showcase the star 2M00044180 to illustrate the synthetic fitting. The plots corresponding to the other three stars are given in Appendix C\footnote{\url{https://doi.org/10.5281/zenodo.13341342}}.

\subsection{Iron\label{Fe}}
Table \ref{table:targetinfo} shows that in three of the four stars, the iron (Fe) abundances obtained in the optical are about \unit[0.10]{dex} lower than the values derived in the NIR. This offset was also observed by \citet{masseron20b} and effects related to departures from LTE have been discussed as an explanation \citep{masseron21}. The magnitude of the offset is in reasonable agreement with the NLTE correction for Fe abundances in stars with similar parameters \citep{bergemann11}. Despite possible NLTE effects, the Fe abundances obtained in the optical for 2M00044180, 2M13472354, and 2M22480199, are in reasonable agreement with those from the NIR \citepalias{brauner23}. \\
\indent In the case of 2M18453994, a larger discrepancy in [Fe/H] is observed. This star is a complex case because of its low temperature and low surface gravity, which result in strong molecular bands that complicate continuum placement, leading to greater uncertainty. Its parameters are also not covered by any grid of NLTE corrections, such as the extended grid from \citet{amarsi16}, making it difficult to estimate the impact of NLTE effects on the Fe lines in this star. A line selection bias compared to the other stars can be ruled out because only three out of the 140 lines were used uniquely for 2M18453994. \\
\indent Although the Fe ion is known to be more robust regarding NLTE effects as it was shown by several studies \citep[e.g.,][]{bergemann11,mashonkina11,lind12}, only one Fe II line at \unit[6456.379]{\angstrom} could be measured. Both values for [FeI/H] and [FeII/H] are consistent within the errors in all four stars. For this reason, we rely on the large number of Fe I lines in the optical (between 140 and 242) to determine the Fe abundance which is used as metallicity proxy, also to keep the analysis consistent regarding the spectral range that we study in this work.

\subsection{Gallium and germanium\label{GaGe}}
We found that the synthesis of the single gallium (Ga) line \unit[4172.042]{\angstrom} was slightly too strong compared to the observed solar spectrum. We changed the log(gf) value to the one used by \citet{sneden03} and achieved a better agreement. Absorption features of Ti II and Fe I on the blueward side and of Fe I on the redward side close to the Ga line complicate the modeling. We modified the log(gf) values of these features to achieve a better match. \\
\indent The sole germanium (Ge) line at \unit[4685.829]{\angstrom} appears to be strong in all stars except for 2M13472354, where it is too weak to yield a reliable measurement. The line is blended by a Co I line, which is not expected to significantly alter the Ge abundance because of its low oscillator strength.

\subsection{Rubidium\label{Rb}}
The two rubidium (Rb) lines used for abundance determination are shown in Fig. \ref{fig:2M00Rb}. The stronger one, Rb I at \unit[7800.250]{\angstrom}, appears on the wing of a strong Si I line. Because the Si abundance is well constrained, this Si I line is not expected to affect the measurement of Rb. In three of the four stars, both Rb I lines are well fitted by the same abundance. In the case of 2M13472354, the spectral region of the Rb I lines is missing. \\
\indent In Table \ref{table:ratios}, we present [Rb/Zr] and [Rb/Sr], which are known to provide information on the neutron source \citep[e.g.,][and references therein]{contursi23} as these elements are produced by the s-process. A negative [Rb/Zr] (and/or [Rb/Sr]) points at the \ce{^{13}C}($\alpha$,n)\ce{^{16}O} reaction which occurs in low-mass ($\sim$1--3 M$_\odot$) AGB stars as the dominant neutron source (see e.g., \citealp[and references therein]{garcia-hernandez06}). We also added to Table \ref{table:ratios} the ratios of the two targets studied by \citet{masseron20b}, 2M13535604+4437076 and 2M22045404-1148287. All ratios are in line with the results from \citet{masseron20b}, who concluded that the s-process contribution occurs at rather low neutron densities.

\subsection{Strontium\label{Sr}} 
\begin{figure}
  \resizebox{\hsize}{!}{\includegraphics{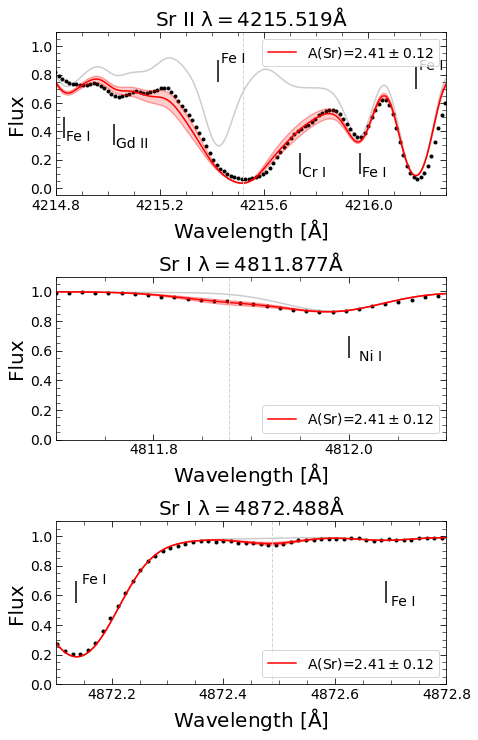}}
  \caption{Same as Fig. \ref{fig:2M00Rb} for 2M00044180 around the two Sr I and one Sr II lines.}
  \label{fig:2M00Sr}
\end{figure}

The strong lines of ionized strontium (Sr), \unit[4077.709]{\angstrom} and \unit[4215.519]{\angstrom}, are traditionally used for abundance determination. The first line is surrounded by transitions of species such as Dy II and Y I. The same is true for the latter, which, apart from having contiguous transitions of Fe I and Cr I, has the \unit[4215.519]{\angstrom} line blended by a Fe I feature in its center, distorting its shape (see Fig. \ref{fig:2M00Sr}). Since the abundance of Fe is well constrained, we used both lines, if available, to derive the Sr abundance. \\
\indent To complement the strong Sr II lines, we also included two weaker Sr I lines in our analysis, both displayed in Fig. \ref{fig:2M00Sr}. One of them, \unit[4811.877]{\angstrom} is located on the blueward wing of a Ni I line. The other one, \unit[4872.488]{\angstrom}, is largely unaffected by blends. \\
\indent To minimize the influence of possible NLTE effects on the abundance results, we focused on accurately fitting the wings of the strong lines, as NLTE effects are known to become visible in the line cores. For the Sr II line \unit[4077.709]{\angstrom}, NLTE effects were examined by \citet{bergemann12sr} and \citet{hansen13}. Using the grid presented by \citet{bergemann12sr}, the correction that needs to be applied on the abundance derived by means of \unit[4077.709]{\angstrom} is negligible in stars with similar parameters as 2M13472354 and 2M22480199.

\subsection{Yttrium and zirconium\label{YZr}}
In the case of yttrium (Y), more lines of the ion are available than for the neutral species. The abundances found for Y I and Y II agree well within errors, although we note that the abundance is slightly higher for Y II. The majority of the used lines are strong and unblended, but show a relatively large abundance spread reflected by the standard deviation. We suspect that this is due to inaccurate atomic data. \\
\indent The abundances of neutral and ionized zirconium (Zr) are in very good agreement. The high number of observed lines (more than ten for each star) with a low line-to-line spread makes the abundance results reliable.

\begin{figure*}
  \resizebox{\hsize}{!}{\includegraphics{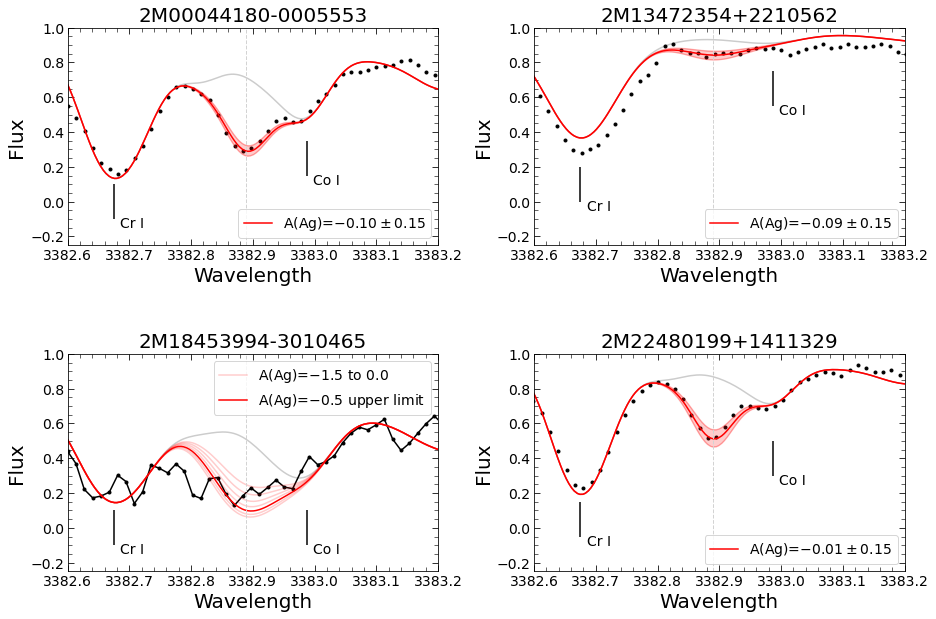}}
  \caption{Observed (black dots) and synthetic spectra around the Ag I line at $\lambda$=\unit[3382.889]{\angstrom} for the four target stars. The synthetic spectra were computed for A(Ag)=$-\infty$ (gray line) and for the final abundance value from Table \ref{table:abuheavy} (red line). The red shaded area corresponds to abundance variations in the amount of the standard deviation. In the case of 2M18453994, several syntheses from A(Ag)=\unit[$-$1.5]{dex} to \unit[0.0]{dex} in steps of \unit[0.25]{dex} are shown to illustrate the choice of the upper limit.}
  \label{fig:Ag_lines}
\end{figure*}

\subsection{Niobium\label{Nb}}
The lines of neutral and ionized niobium (Nb) are mainly located in the crowded blue region of the spectrum. This means that these Nb lines are immersed within blends from multiple species and are often affected by continuum mismatches. The measurements for 2M00044180 and 2M18453994 were only possible after a continuum adjustments and an improvement of the log(gf) value of a blending Fe I line close to Nb I \unit[4168.120]{\angstrom}. The measurements are quite uncertain, as reflected by the large line-to-line abundance spread. For 2M13472354 and 2M22480199, only conservative upper limit estimates were possible due to noise or severe continuum mismatches.

\subsection{Molybdenum\label{Mo}}
To measure molybdenum (Mo), we identified a few suitable lines clearly detectable free from major blends. Most of them are located in the red part of the spectrum (\unit[>5500]{\angstrom}). In the case of 2M13472354, the only line available is \unit[3864.103]{\angstrom}, which is also free of blends and has an excitation potential and oscillator strength concordant with \citet{sneden03} and \citet{ernandes23}. Hence, we declare the measurements of Mo to be reliable.

\subsection{Ruthenium, rhodium, palladium\label{RuRhPd}}
The lines of ruthenium (Ru), rhodium (Rh), and palladium (Pd) are mostly located in the blue region of the spectrum, which causes the same problems as for Nb (see Sect. \ref{Nb}), making it necessary to adjust the continuum around some lines by eye. \\
\indent For Ru, the cleanest lines are the ones located at \unit[4709.482]{\angstrom} and \unit[4869.153]{\angstrom}. After adjusting the continuum of the blue lines, abundances obtained from the lines of both regions are consistent. \\
\indent For Rh, the lines used for abundance measurements are \unit[3396.819]{\angstrom} and \unit[3692.358], and in the case of 2M22480199, we also used \unit[3583.092]{\angstrom}. The additional Rh lines given in the line list are used to constrain a consistent upper limit. \\
\indent For Pd, the only line without blend is \unit[3404.579]{\angstrom}, but we found that this line appears to be too strong in the solar spectrum. Thus, we changed log(gf) to a lower value. In the region of this line, the continuum had to be adjusted by eye. The other two lines are affected by blends and were used for upper limit estimations in the cases of 2M13472354 and 2M18453994.

\begin{figure*}
  \resizebox{\hsize}{!}{\includegraphics{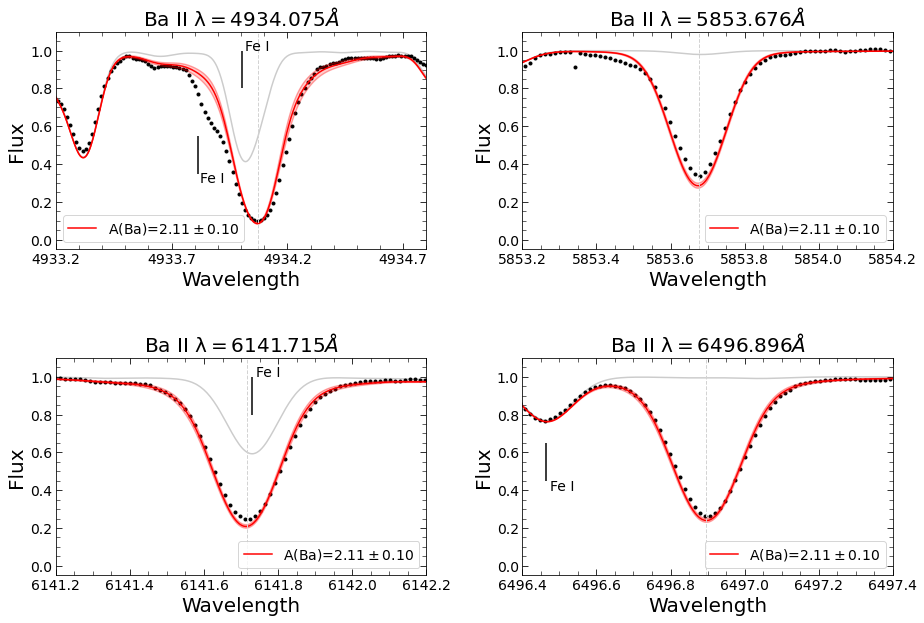}}
  \caption{Same as Fig. \ref{fig:2M00Rb} for 2M00044180 around the four Ba II lines}
  \label{fig:2M00Ba}
\end{figure*}

\subsection{Silver\label{Ag}}
Silver (Ag) is an element rarely studied in the literature and only a few measurements exist (e.g., \citealp{hansen11}). For this reason and because of its strong importance regarding i-process model disambiguation (see Sec. \ref{subsec: iprocess_1}), its abundance determination in all four P-rich stars is highlighted in Fig. \ref{fig:Ag_lines}. The single Ag I line at \unit[3382.889]{\angstrom} is located in the close vicinity of a strong Cr I line and is affected by a Co I blend on its redward wing. We changed the log(gf) value of the Co I line to optimize the measurement of the Ag I line. Nevertheless, we note that the adopted log(gf) produces a too strong line in the solar spectrum. Given that the P-rich stars are not the same stellar type as the Sun, this might be due to 3D or NLTE effects. Apart from this general fix, in the case of 2M13472354, we also adopted a slightly higher Co abundance (A(Co) $=$ 3.72 instead of the obtained average A(Co) $=$ 3.56) to improve the fit to this specific line. Except for 2M18453994, where only an upper limit was estimated due to the strong noise, the Ag I line is well fitted in the other three stars by the synthetic spectra corresponding to the respective abundances given in each panel of Fig. \ref{fig:Ag_lines}.

\subsection{Cadmium, indium, tin, cesium\label{CdInSnCs}}
The most prominent cadmium (Cd) feature at \unit[4799.914]{\angstrom} is located on the wing of a feature produced by transitions of Ti I, V I, Ni I, and Fe I. These multiple components complicate the fitting of the blend and it becomes difficult to obtain a reliable measurement of Cd. Moreover, the Cd I line is also blended in its center by a Ca II feature. The log(gf) of the Fe I line on the redward side close to the Cd I line was adjusted for a better estimation of the upper limits.  \\
\indent For indium (In), only the line \unit[4511.307]{\angstrom} is susceptible to changes in abundance. Given the presence of several Fe features at or around the In line center, together with blends of Cr I and V I, we estimated an upper limit for the In abundance. \\
\indent For tin (Sn), two lines in the blue part of the spectrum are available, but both are severely blended and affected by continuum mismatches. Adjustments of the continuum and changes to the log(gf) of a close Ce II feature were applied to ease the estimation of upper limits for Sn. \\
\indent A weak line of cesium (Cs) is present at \unit[8521.131]{\angstrom}. We estimated an upper limit for Cs because multiple blends are affecting this line. Other than for Cd, In, and Sn, the upper limits obtained for Cs are less restrictive.

\begin{figure*}
  \resizebox{\hsize}{!}{\includegraphics{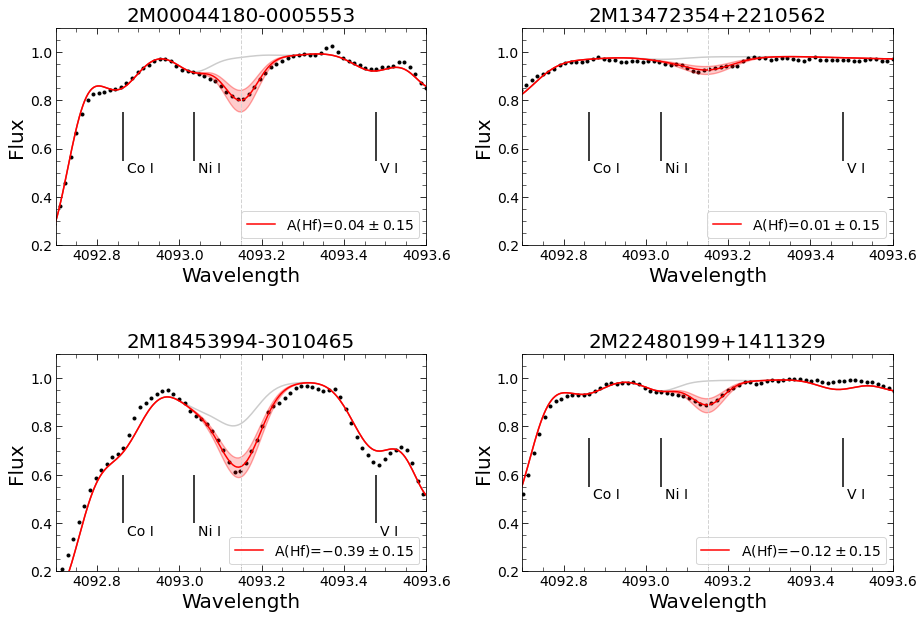}}
  \caption{Observed (black dots) and synthetic spectra around the Hf I line at $\lambda$=\unit[4093.150]{\angstrom} for the four target stars. The synthetic spectra were computed for A(Hf)=$-\infty$ (gray line) and for the final abundance value from Table \ref{table:abuheavy} (red line). The red shaded area corresponds to abundance variations in the amount of the standard deviation.}
  \label{fig:Hf_lines}
\end{figure*}

\begin{figure*}
  \resizebox{\hsize}{!}{\includegraphics{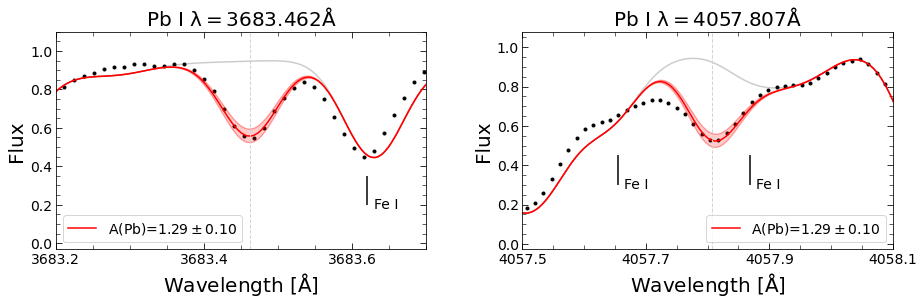}}
  \caption{Same as Fig. \ref{fig:2M00Rb} for 2M00044180 around the two Pb I lines}
  \label{fig:2M00Pb}
\end{figure*}

\subsection{Barium\label{Ba}}
For the barium (Ba) abundance, we rely on four strong Ba II lines (see Fig. \ref{fig:2M00Ba}). Two of them, \unit[4934.075]{\angstrom} and \unit[6141.715]{\angstrom}, are contaminated by Fe I. The remaining two, \unit[5853.676]{\angstrom} and \unit[6496.896]{\angstrom}, are free of blends. The traditionally used resonance line \unit[4554]{\angstrom} is not measured because it falls in the CCD gap. \\
\indent Similar to Sr (see Sect. \ref{Sr}), NLTE effects have to be considered when discussing absolute Ba abundances. To alleviate their influence, we prioritize a high quality fit of the wing because NLTE effects become visible in the cores of such strong lines. Correction grids for Ba II lines are provided by \citet{korotin15} and \citet{mashonkina19}, but the parameters of the four target stars are not covered by those grids. Therefore, we did not apply any NLTE correction. A recent study about departures from LTE by \citet{amarsi20} found a median correction for the two lines \unit[5853.679]{\angstrom} and \unit[6496.904]{\angstrom} for a type of star with parameters similar to the ones of our targets. Applying this correction would reduce the abundance by approximately \unit[0.12]{dex} and \unit[0.18]{dex}, respectively. This might improve the fit in the core, but without affecting the fit to the wing or the overall abundance significantly. In other words, we do not expect that including NLTE corrections changes the strong Ba enhancement significantly. We add that the NLTE corrections offered by the references given above are made for chemical compositions that are distinct from P-rich stars, meaning that a straightforward application is not possible. \\
\indent To probe the impact of a possible inaccurate isotopic ratio, we calculated the abundance with isotopic mixtures different from solar. In particular, we performed two calculations: assuming a composition of only \ce{^138Ba} and another one with only \ce{^137Ba}. We found that the \unit[4934.075]{\angstrom} line is the most sensitive to changes in the isotopic ratio. Nevertheless, the overall abundance is not altered with such extreme changes in the ratio, meaning that if our isotopic ratio was inaccurate, this would not affect significantly the Ba abundance beyond the error bar. Thus, for the final abundance calculation, we fixed a solar mixture. \\
\indent In Table \ref{table:ratios}, we also give the [Ba/La] abundance ratio. The positive and, in the case of 2M13472354 and 2M18453994, high ratio manifests the overabundance of Ba compared to other second peak s-process elements, such as lanthanum.

\subsection{Lanthanum and lanthanides\label{La}}
The ionized species of lanthanum (La), the lanthanides cerium (Ce), praseodymium (Pr), neodymium (Nd), and samarium (Sm) provide a significant number of strong and unblended lines. We therefore rate the results as reliable. Fewer lines are available for europium (Eu) and dysprosium (Dy), but their line-to-line abundances do not show any remarkable scatter. Hence, we also consider their abundance results as reliable. Lines of ionized lutetium (Lu) are heavily blended and we used the feature around \unit[3507.4]{\angstrom}, in addition to the lines \unit[5476.675]{\angstrom} and \unit[6221.890]{\angstrom}, to estimate an upper limit for the abundance of Lu.

\subsection{Hafnium\label{Hf}}
The hafnium (Hf) abundance was derived with the single line \unit[4093.150]{\angstrom}, displayed in Fig. \ref{fig:Hf_lines}, together with its fit in each star. We applied some changes to the log(gf) values of nearby features. The most important one is the Ni I blend on the bluesided wing, which was reduced from log(gf)=--0.925 to log(gf)=--1.125 to minimize its influence on the Hf abundance. The slight enhancement of \unit[0.34]{dex} on average (Table \ref{table:abuheavy}) is in contrast to the high value of \unit[1.24 $\pm$ 0.20]{dex} found by \citet{masseron20b} in one of their targets.

\subsection{Lead\label{Pb}}
Lead (Pb) forms part of the third peak s-process elements. In a previous study \citep{masseron20b}, it was detected in one of the two P-rich stars considered. \citet{masseron20b} found a strong enhancement of [Pb/Fe] = \unit[1.17$\pm$ 0.20]{dex}, derived from the \unit[4057.807]{\angstrom} line in one P-rich star. We confirm the enhancement, but in a more moderate amount. Overall, we found Pb to be only slightly enhanced, with an abundance of \unit[0.46]{dex} on average. \\
\indent In the wavelength range covered by the spectra, three Pb I lines are available, but one of them is severely blended. We therefore only used \unit[3683.462]{\angstrom} and \unit[4057.807]{\angstrom}. In the case of \unit[3683.462]{\angstrom}, we adopted the log(gf) of the Pb I line to the value used by \citet{sneden03} and \citet{ernandes23}, and also changed the log(gf) of the nearby transitions of Fe I. For \unit[4057.807]{\angstrom}, we only changed the log(gf) of the nearby species but kept the log(gf) of the Pb I line as provided by the VALD. Both lines could be fitted by the same abundance, although the quality of the fit is slightly better for the cleaner \unit[3683.462]{\angstrom} line. Both Pb I lines are shown in Fig. \ref{fig:2M00Pb}, together with their corresponding fits.

\FloatBarrier
\clearpage

\onecolumn

\section{Additional figures\label{AppendixFigures}}

\begin{figure*}[hbt!]
  \includegraphics[width=0.5\textwidth]{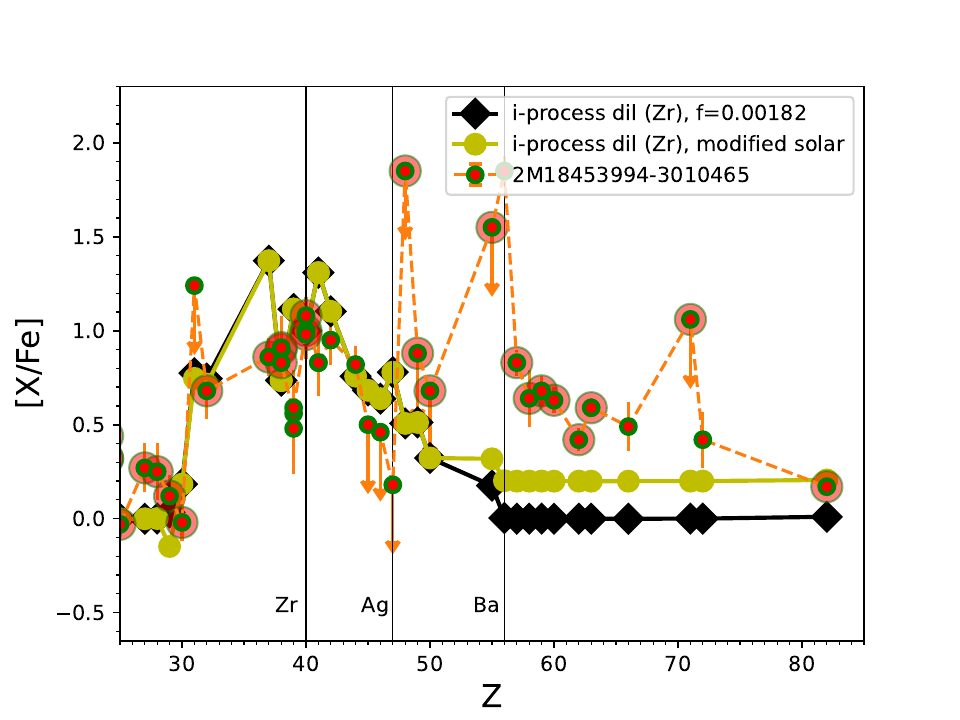} \includegraphics[width=0.5\textwidth]{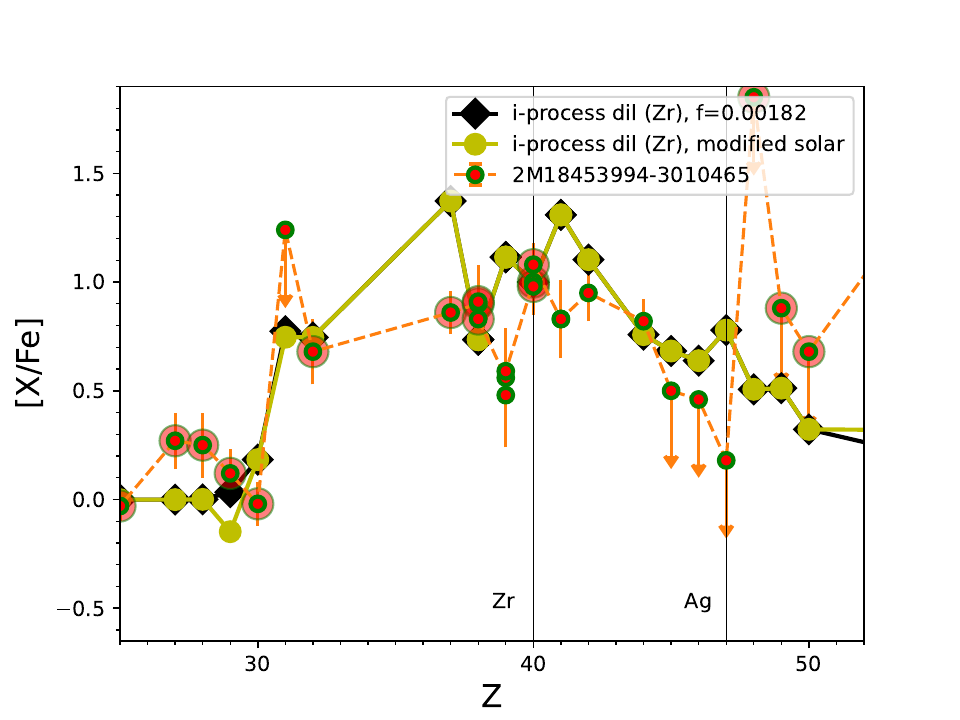}
  \includegraphics[width=0.5\textwidth]{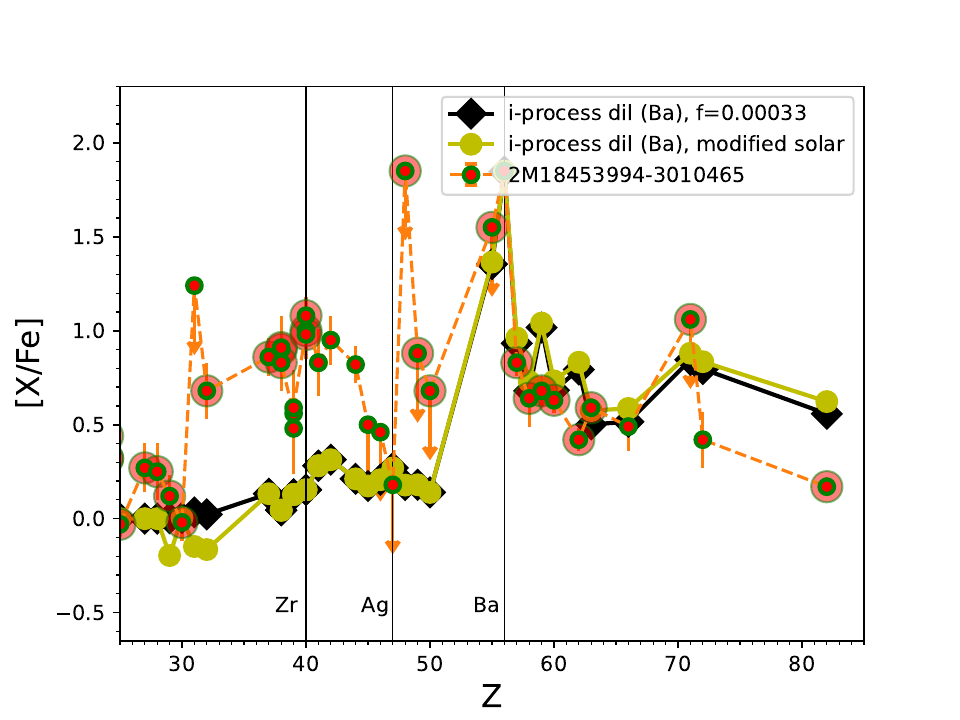}
  \includegraphics[width=0.5\textwidth]{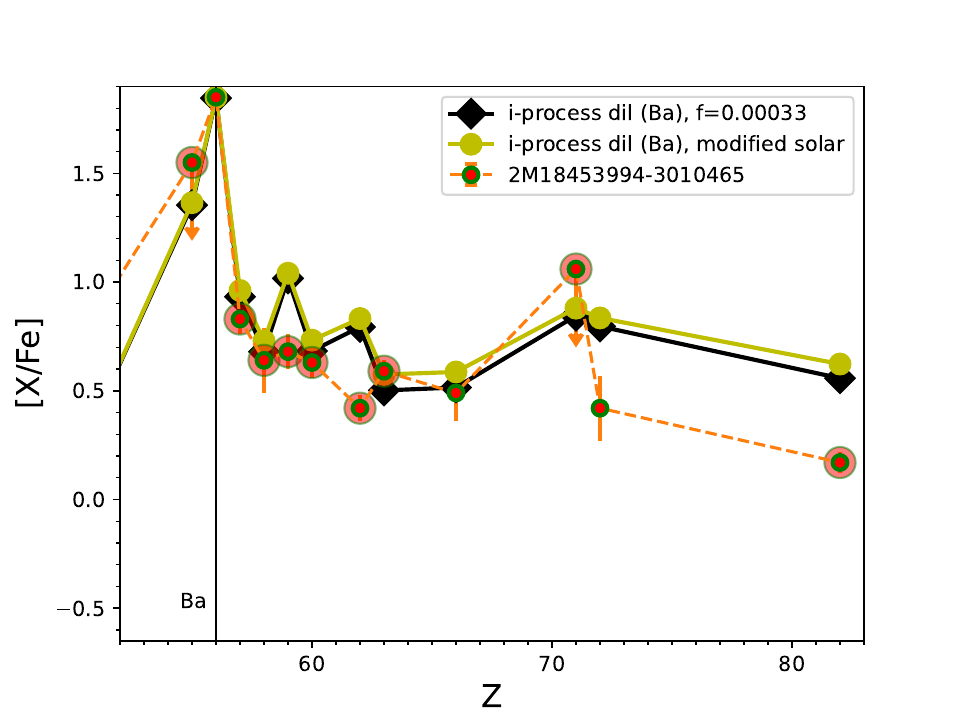}
  \includegraphics[width=0.5\textwidth]{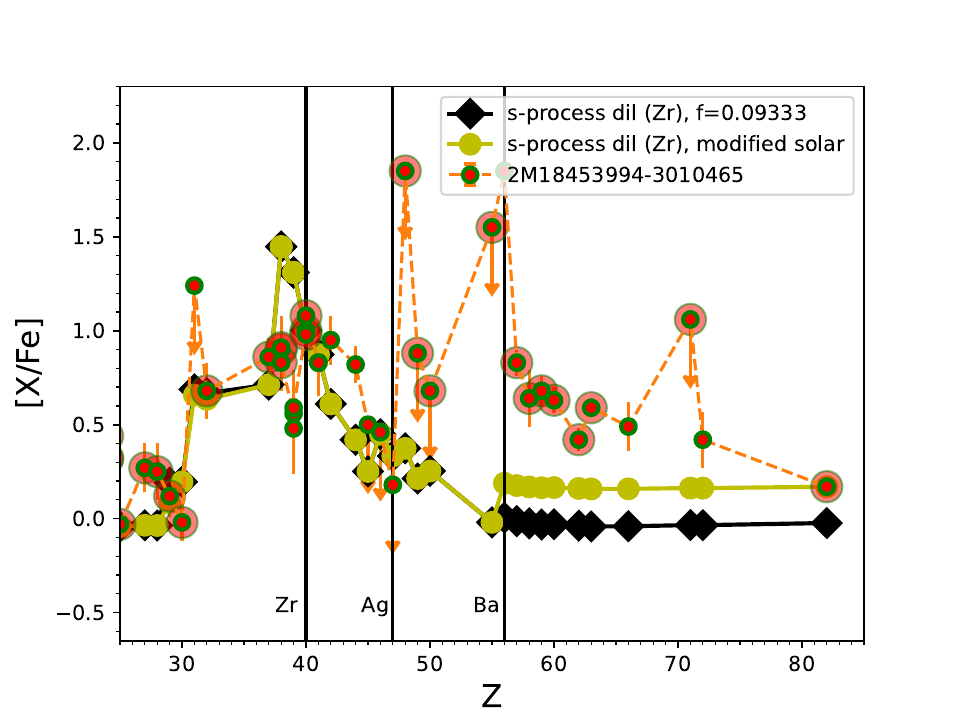}
  \includegraphics[width=0.5\textwidth]{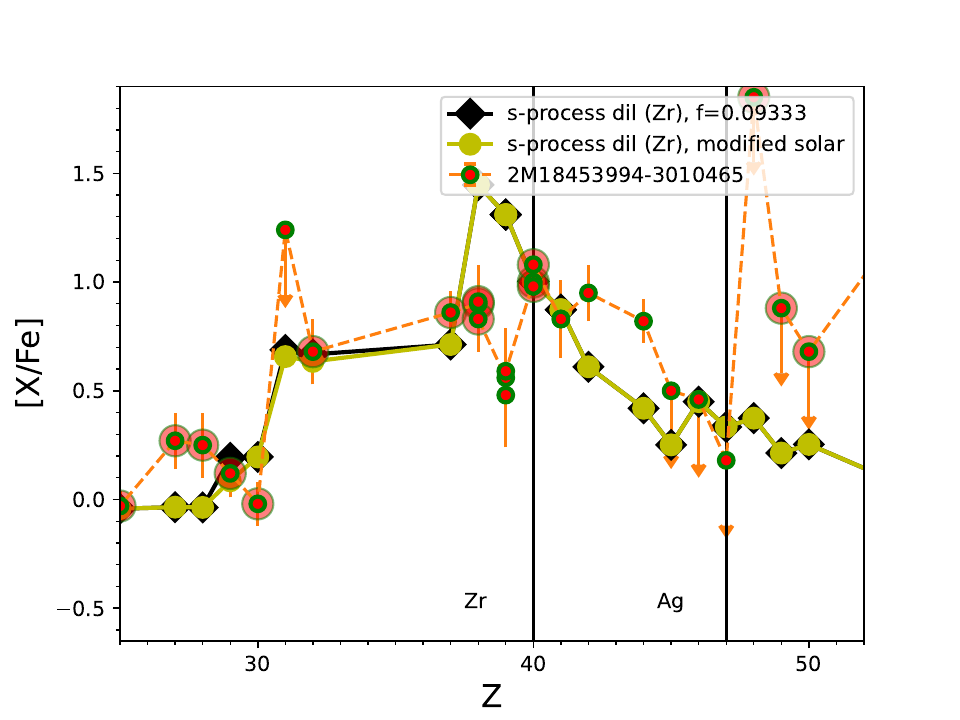}
  \caption{Same as Fig. \ref{fig: scenario1_2M22480199+1411329}, but for 2M18453994.   
  }
  \label{fig: scenario1_2M18453994-3010465}
\end{figure*}

\begin{figure*}
  \includegraphics[width=0.5\textwidth]{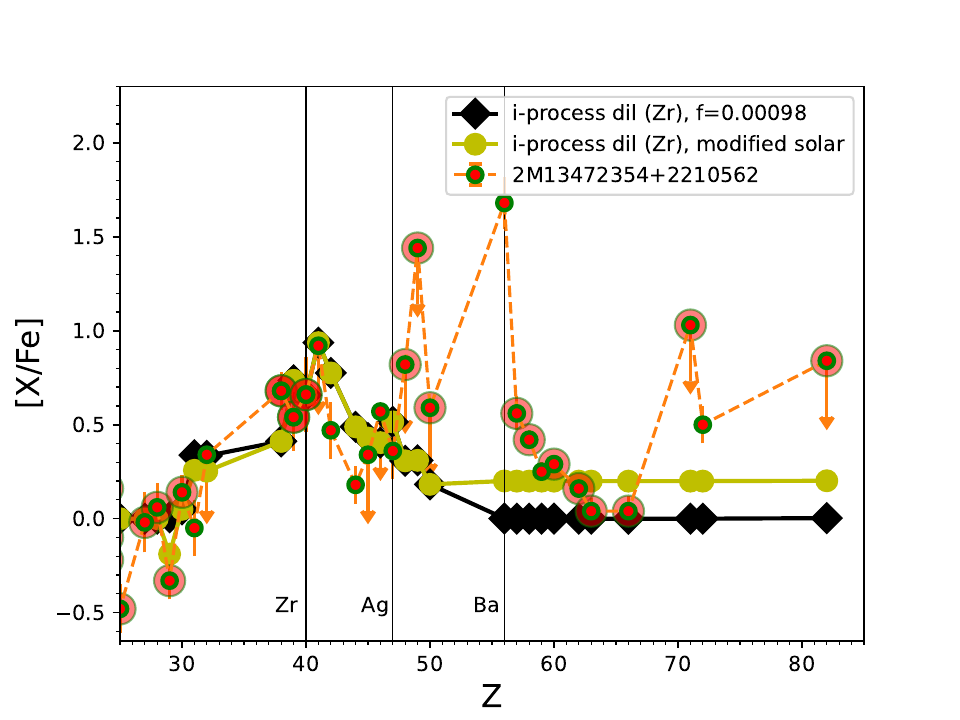}
  \includegraphics[width=0.5\textwidth]{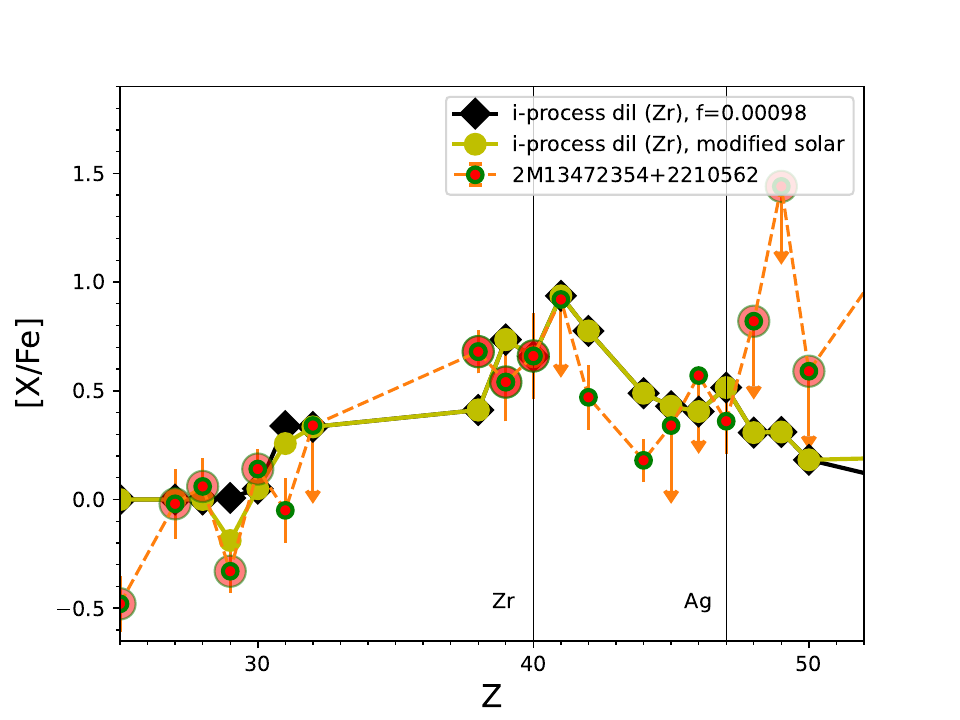}
  \includegraphics[width=0.5\textwidth]{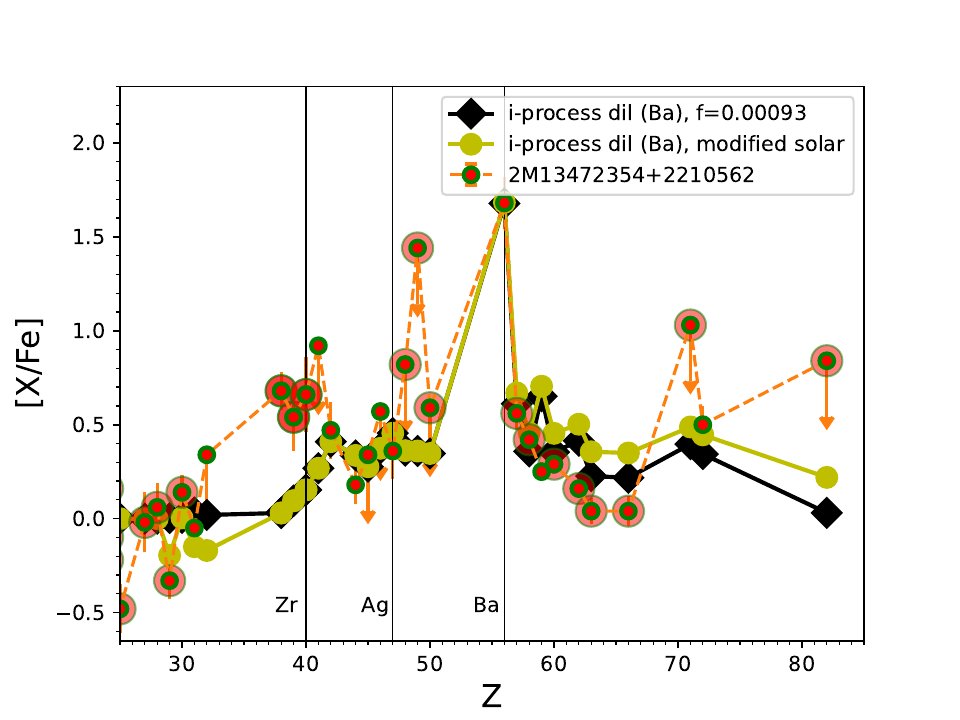}
  \includegraphics[width=0.5\textwidth]{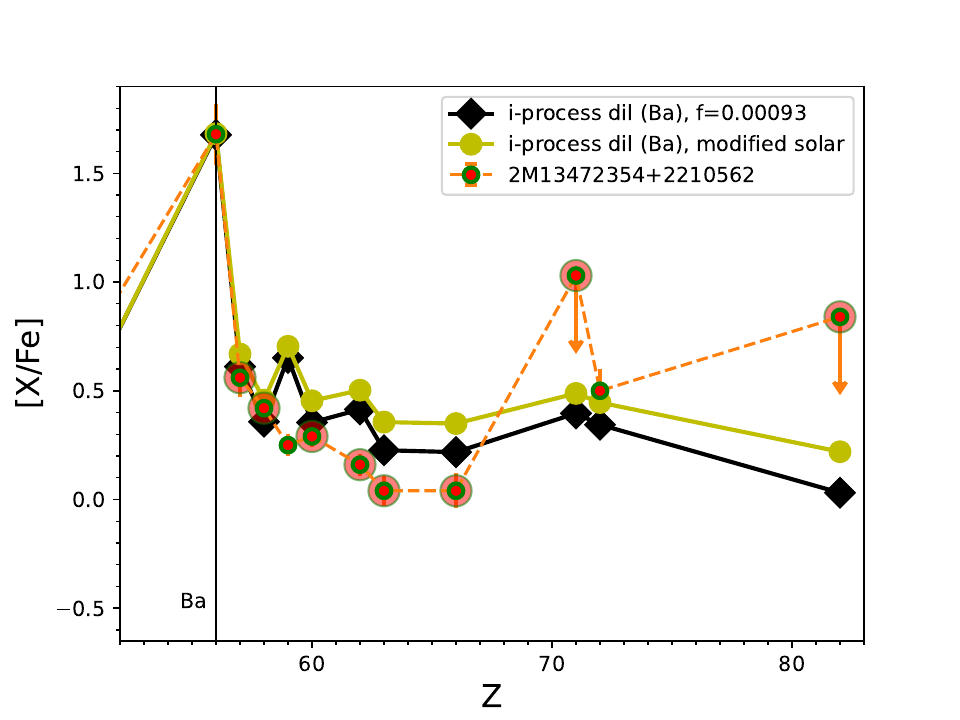}
  \includegraphics[width=0.5\textwidth]{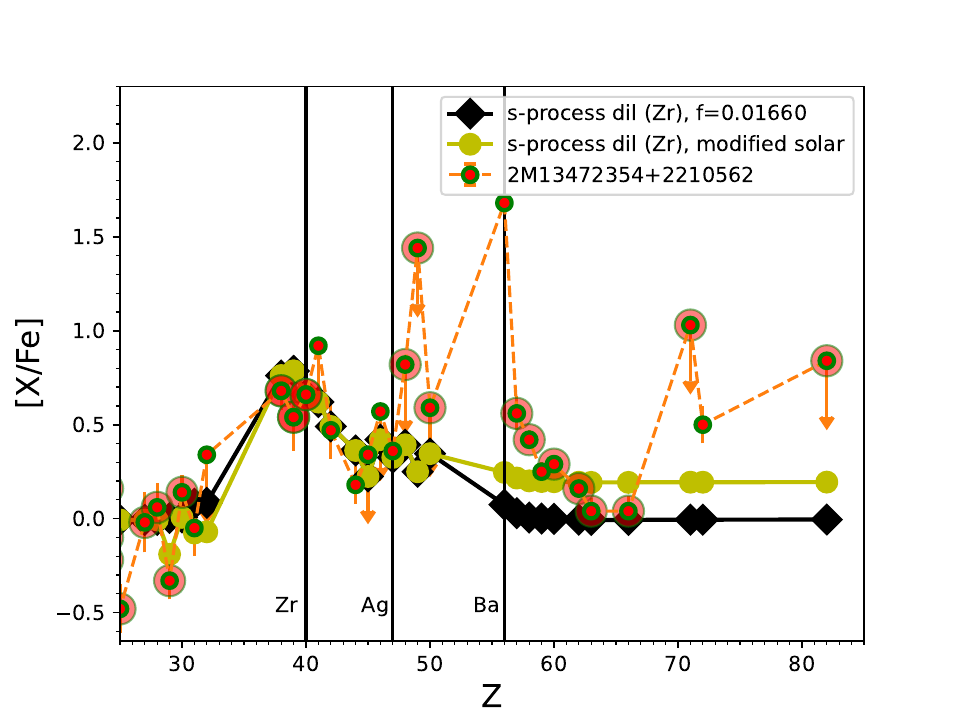}
  \includegraphics[width=0.5\textwidth]{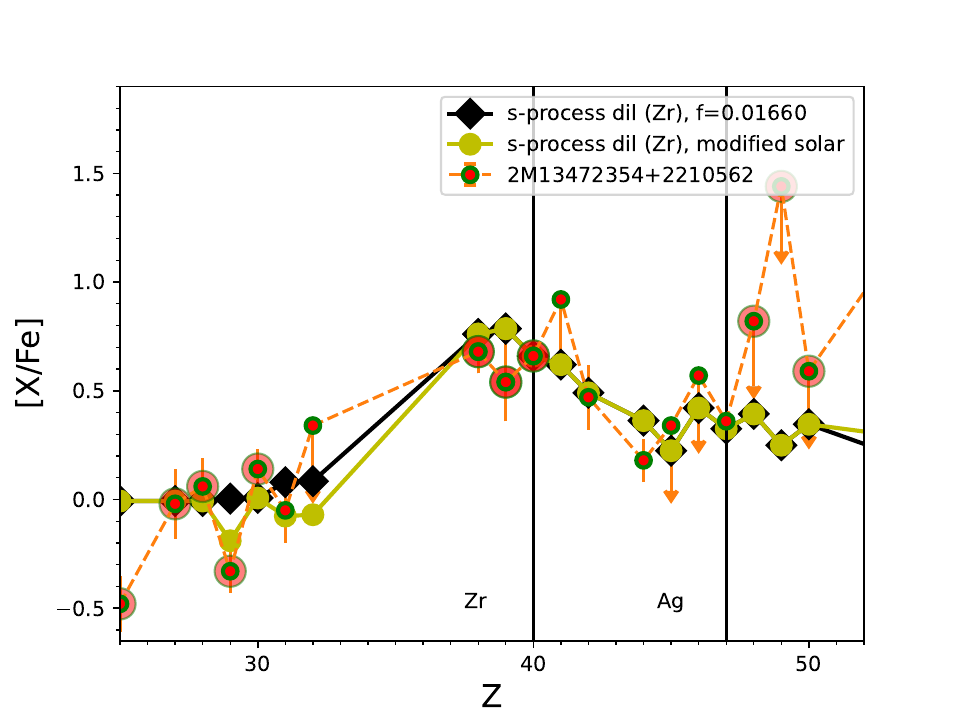}
  \caption{Same as Fig. \ref{fig: scenario1_2M22480199+1411329}, but for 2M13472354.   
  }
  \label{fig: scenario1_2M13472354+2210562}
\end{figure*}

\begin{figure*}
  \includegraphics[width=0.5\textwidth]{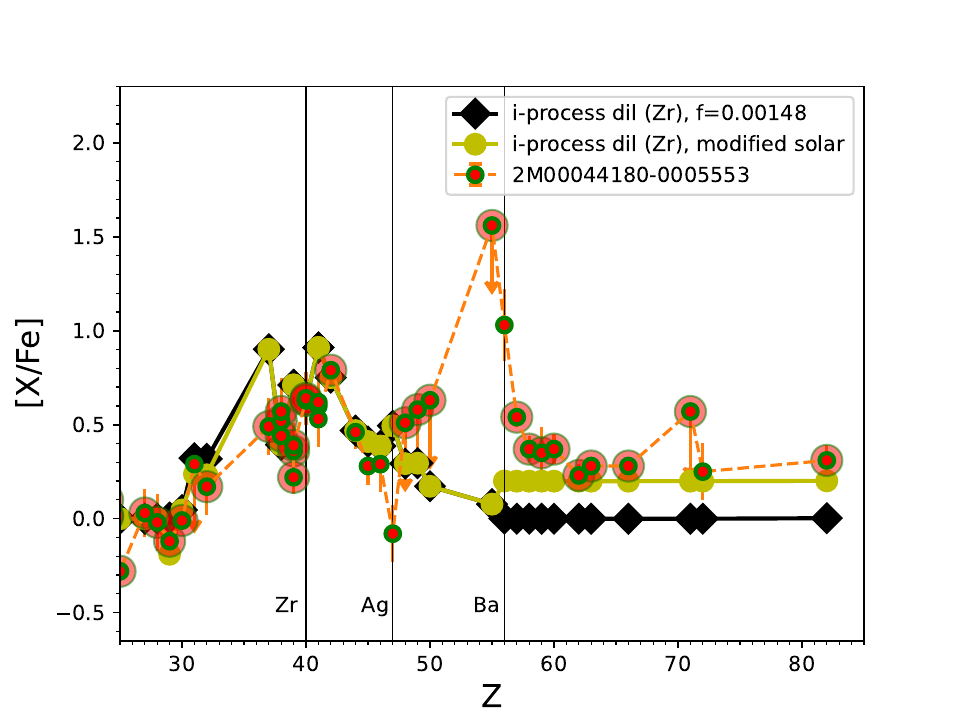}
  \includegraphics[width=0.5\textwidth]{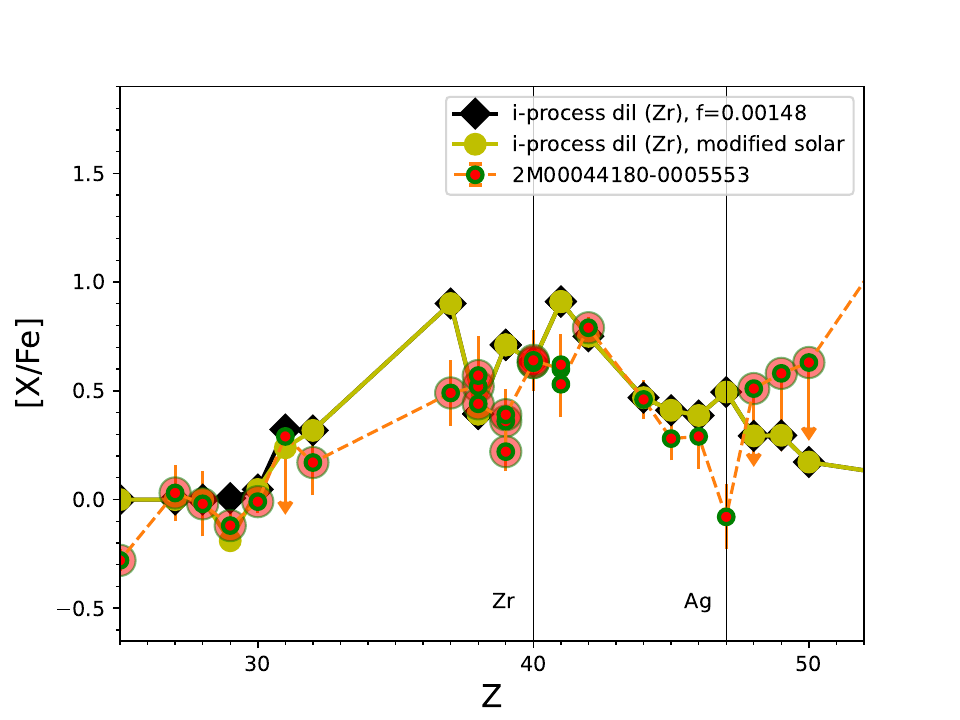}
  \includegraphics[width=0.5\textwidth]{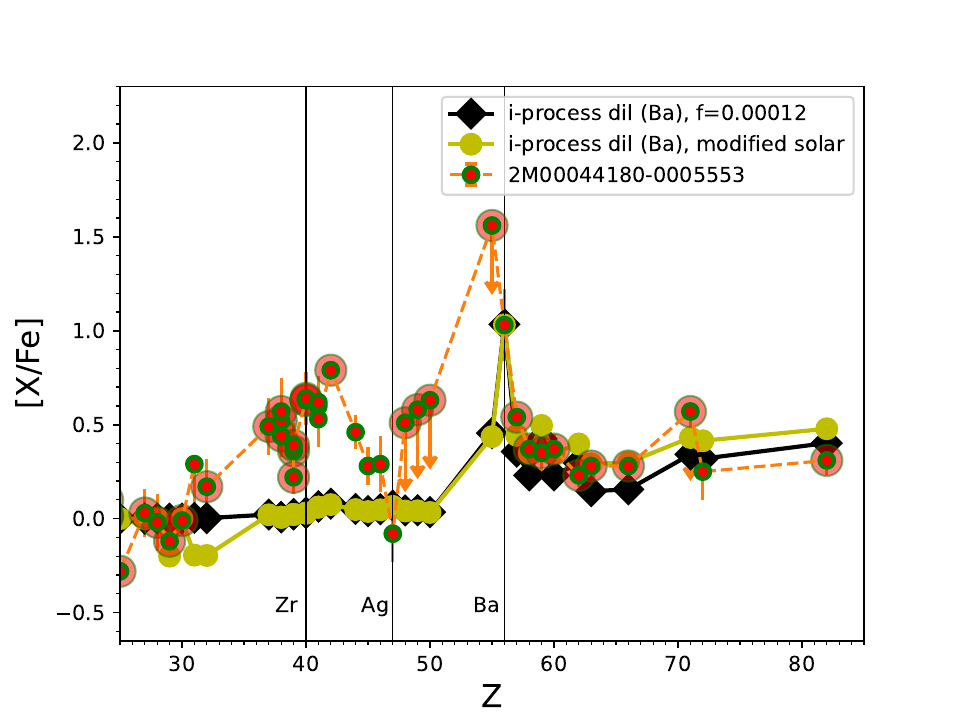}
  \includegraphics[width=0.5\textwidth]{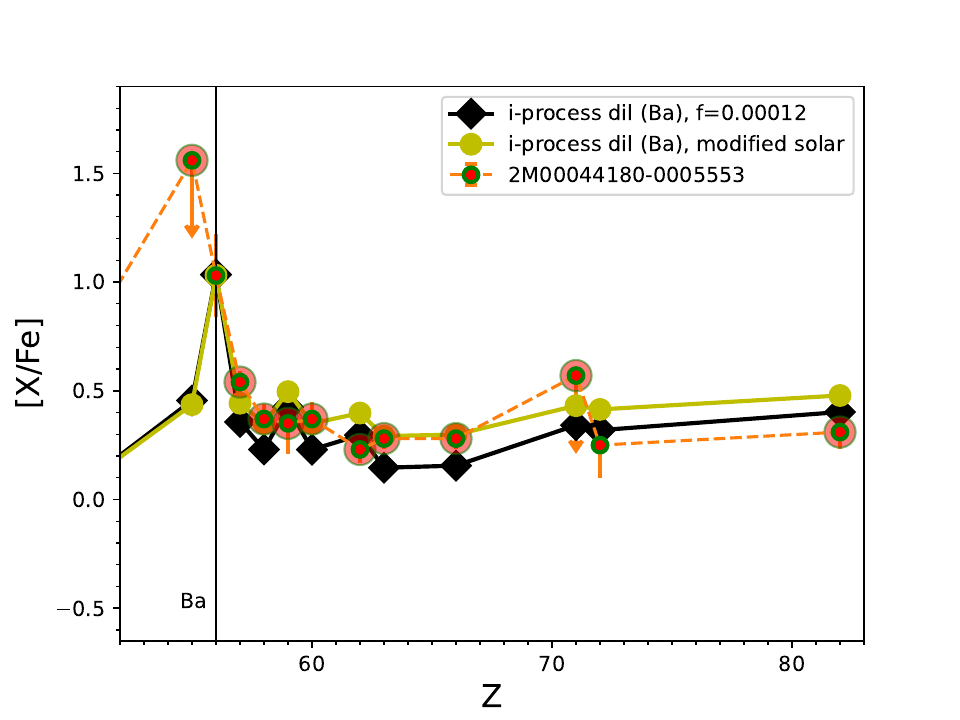}
  \includegraphics[width=0.5\textwidth]{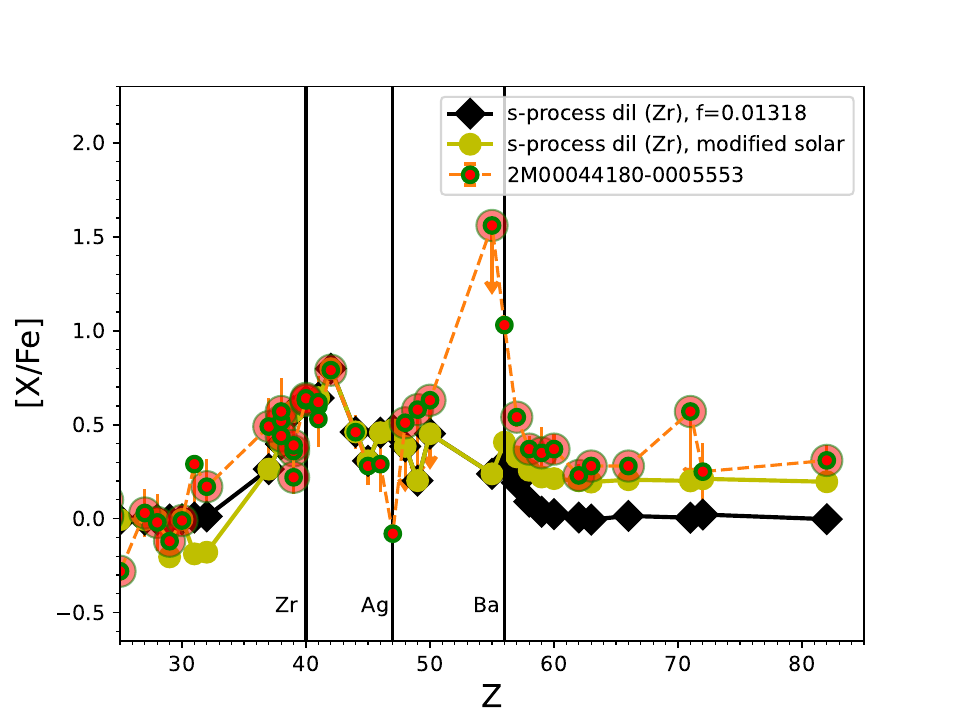}
  \includegraphics[width=0.5\textwidth]{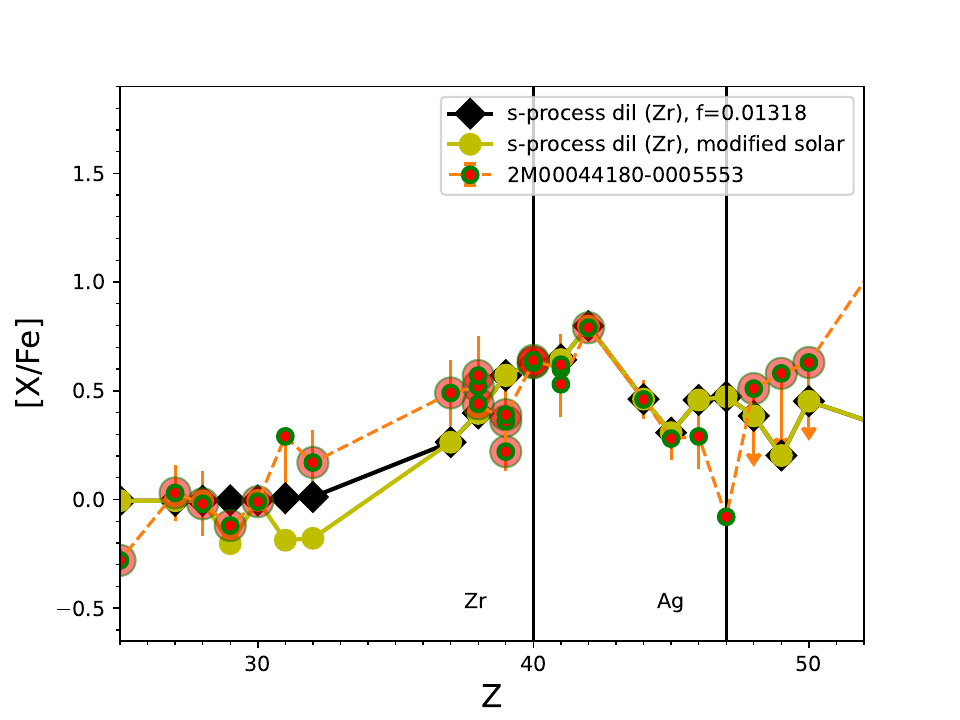}
  \caption{Same as Fig. \ref{fig: scenario1_2M22480199+1411329}, but for 2M00044180.   
  }
  \label{fig: scenario1_2M00044180-0005553}
\end{figure*}

\begin{figure*}
  \includegraphics[width=0.5\textwidth]{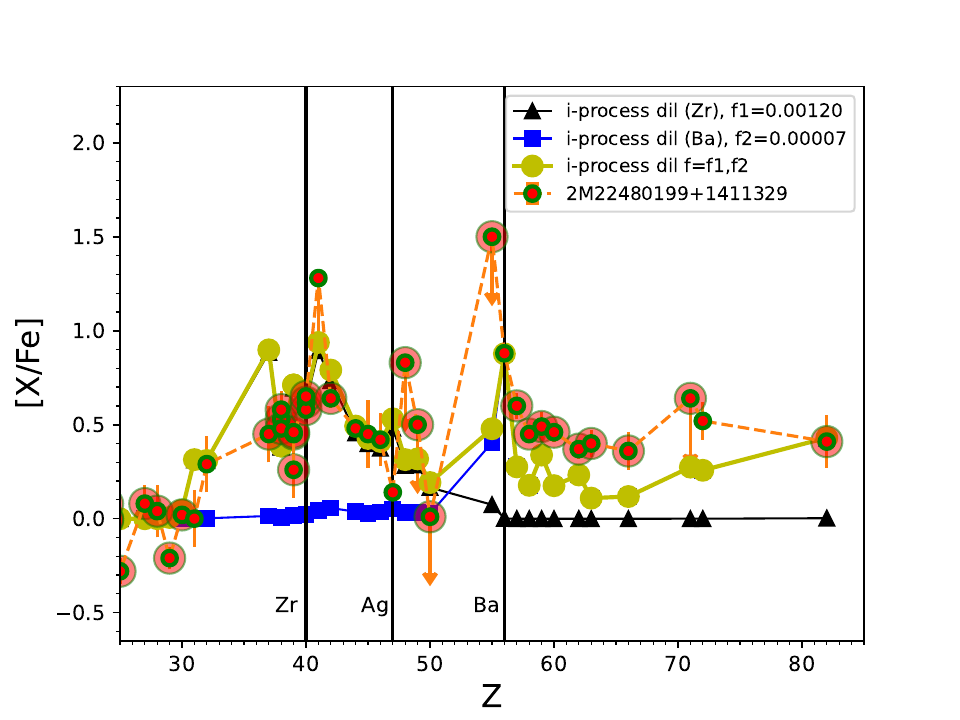}
  \includegraphics[width=0.5\textwidth]{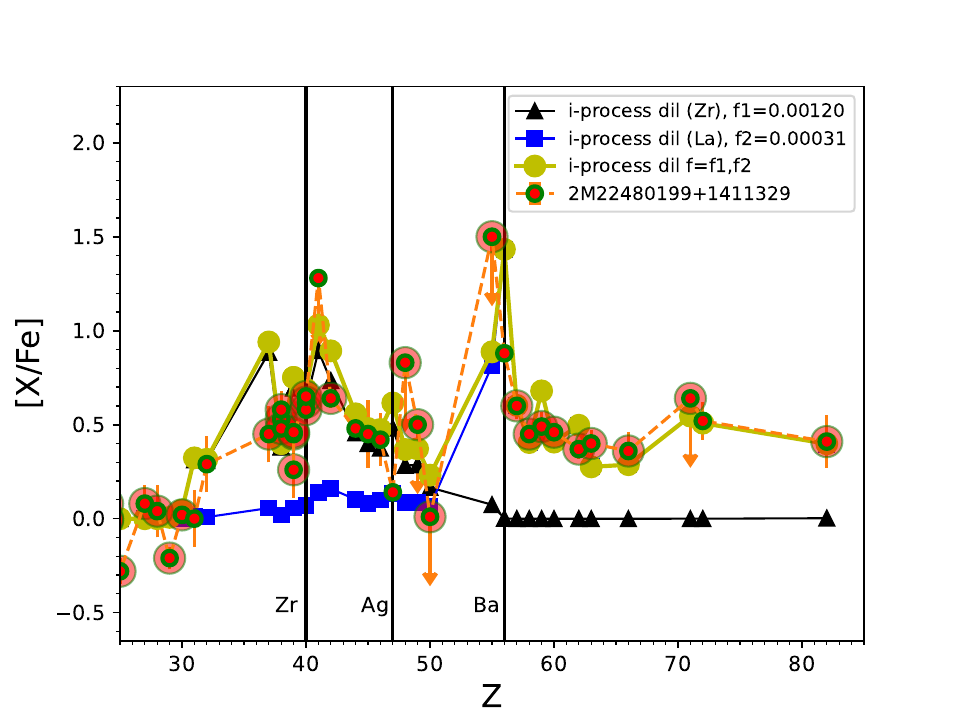}
  \caption{Elemental abundances beyond Fe of 2M22480199 compared with the case 2 models, that is, the two i-process components: one peaked at the neutron-shell closure N=50 (with the observed Zr abundance as a reference to derive the dilution factor f1, blue-squares line) and the second component peaked at N=82 (with Ba as a reference to derive f2 in the left panel, and La in the right panel, black-triangles line), and with the abundance pattern resulting from the sum of the two components (gold-circles line). The elements with the most reliable observational results (measurements or upper limits) are highlighted with large orange circles. 
  }
  \label{fig: scenario2_2M22480199+1411329}
\end{figure*}

\begin{figure*}
  \includegraphics[width=0.5\textwidth]{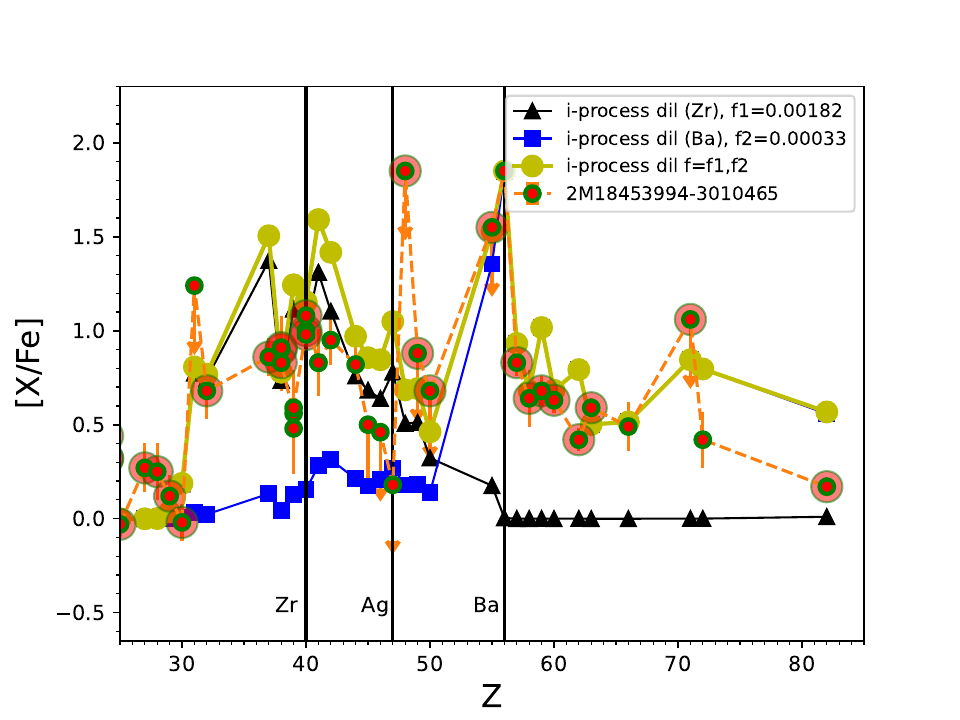}
  \includegraphics[width=0.5\textwidth]{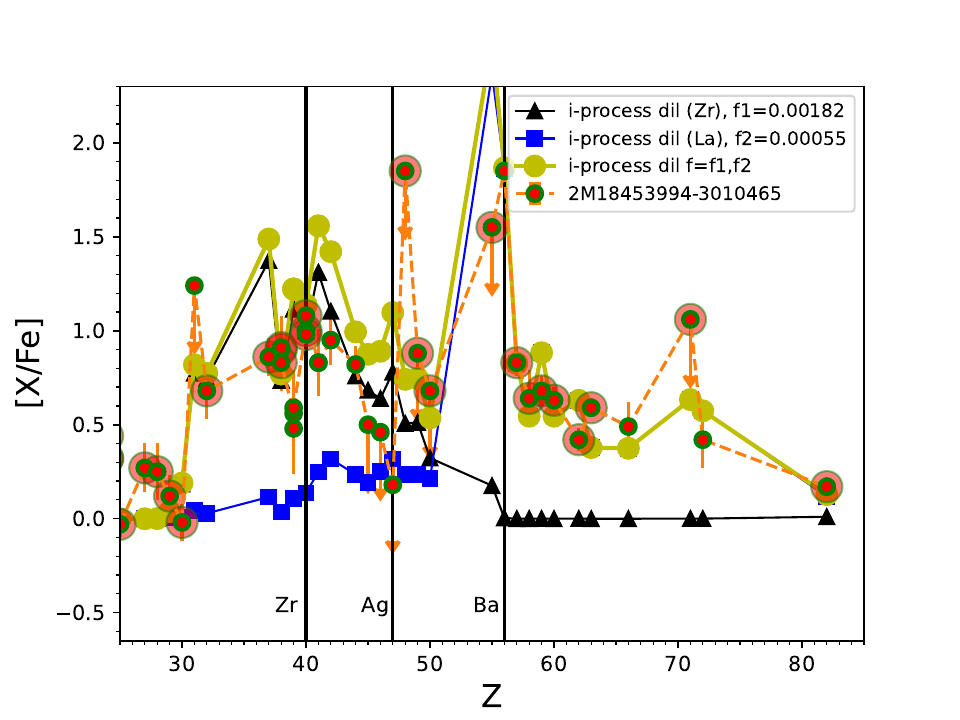}
  \caption{Same as Fig. \ref{fig: scenario2_2M22480199+1411329}, but for 2M18453994. 
  }
  \label{fig: scenario2_2M18453994-3010465}
\end{figure*}

\begin{figure*}
  \includegraphics[width=0.5\textwidth]{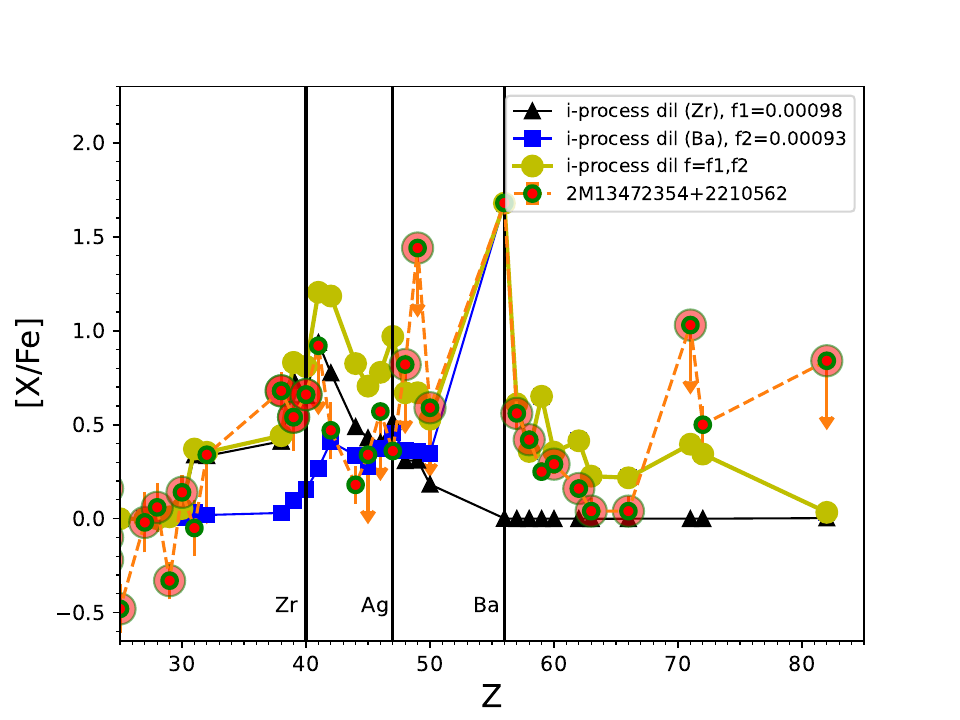}
  \includegraphics[width=0.5\textwidth]{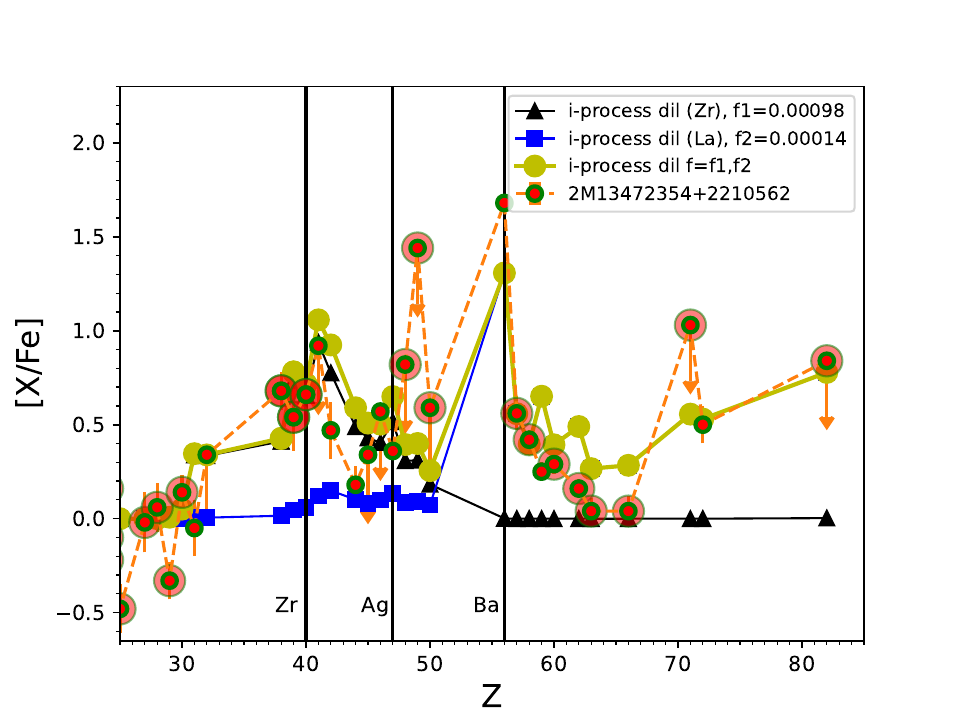}
  \caption{Same as Fig. \ref{fig: scenario2_2M22480199+1411329}, but for 2M13472354. 
  }
  \label{fig: scenario2_2M13472354+2210562}
\end{figure*}

\begin{figure*}
  \includegraphics[width=0.5\textwidth]{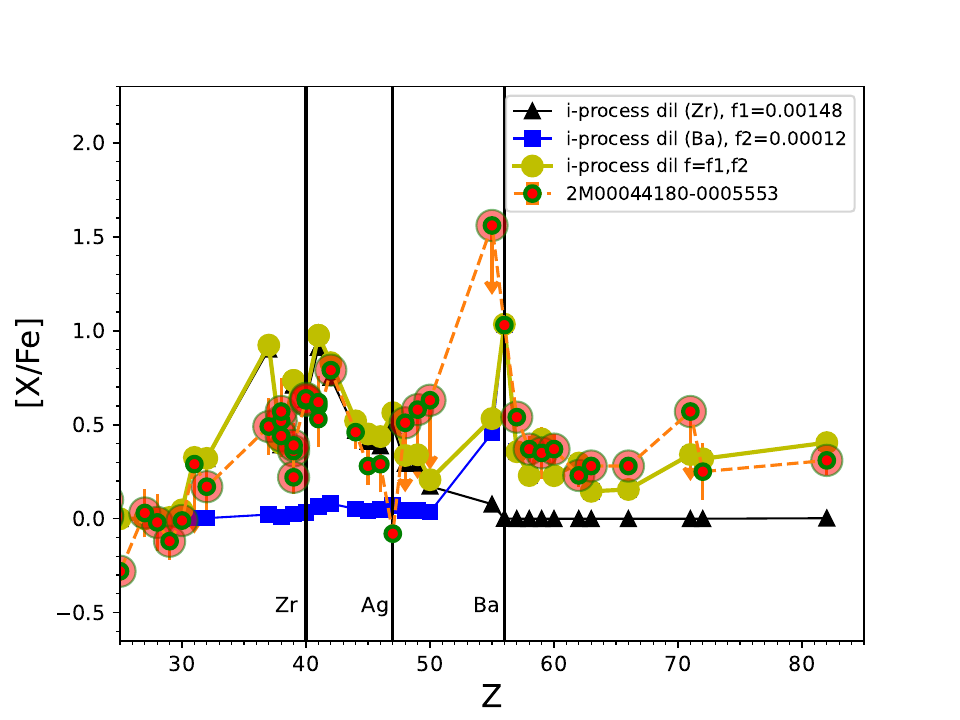}
  \includegraphics[width=0.5\textwidth]{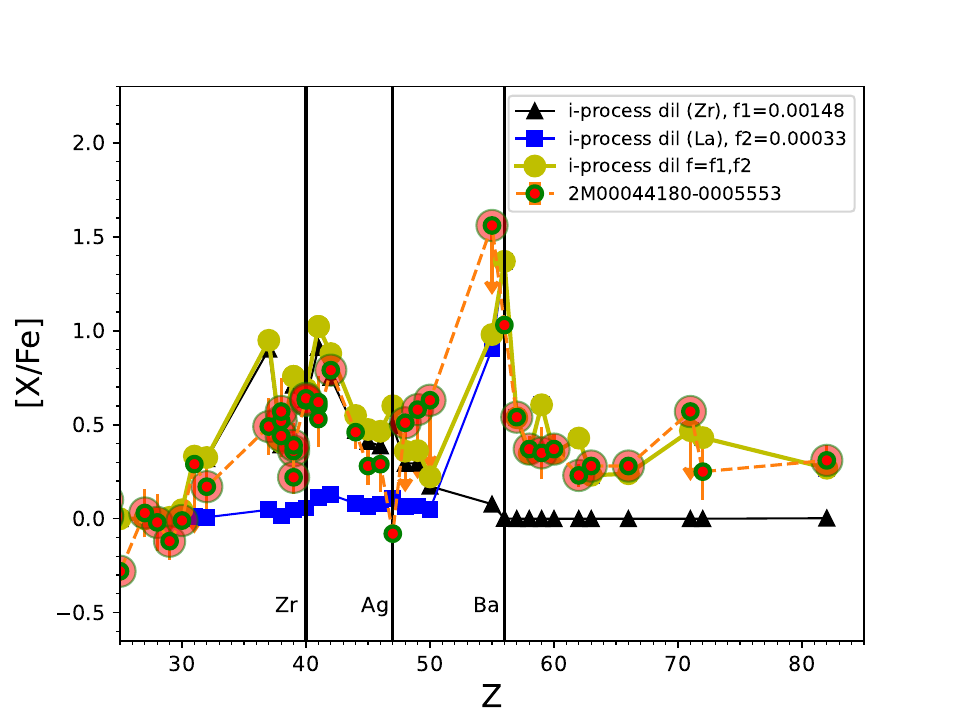}
  \caption{Same as Fig. \ref{fig: scenario2_2M22480199+1411329}, but for 2M00044180. 
  }
  \label{fig: scenario2_2M00044180-0005553}
\end{figure*}

\begin{figure*}
  \includegraphics[width=0.5\textwidth]{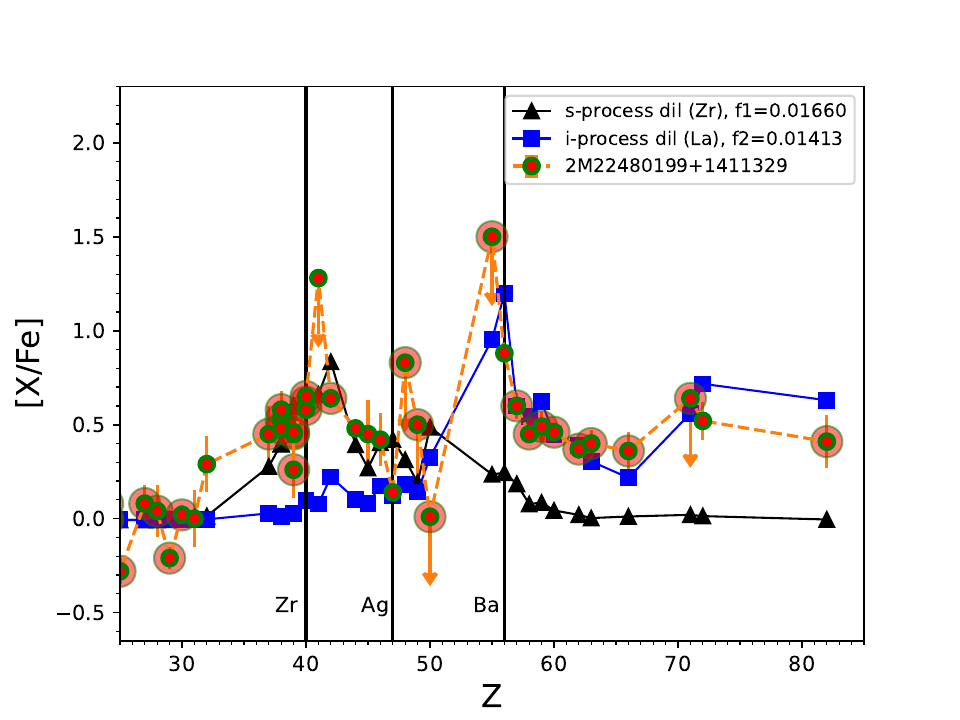}
  \includegraphics[width=0.5\textwidth]{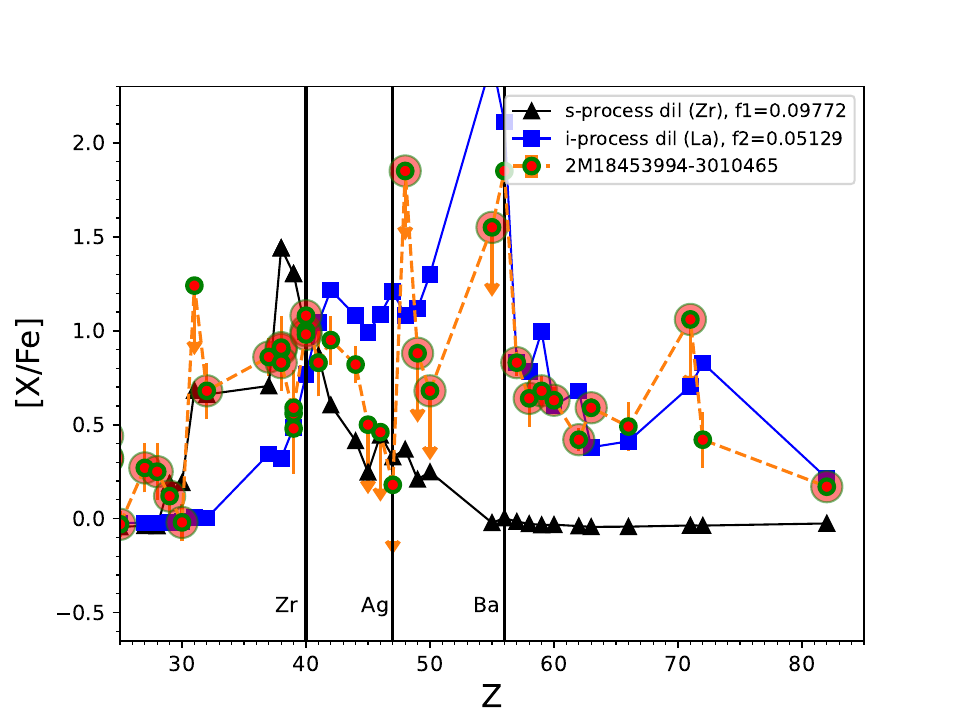}
  \includegraphics[width=0.5\textwidth]{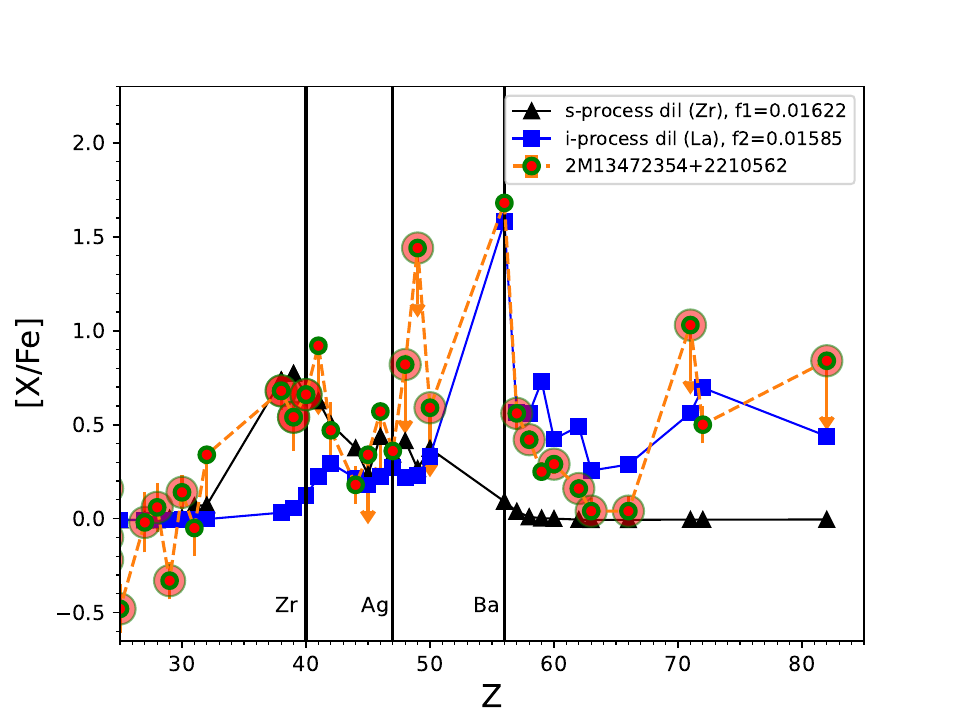}
  \includegraphics[width=0.5\textwidth]{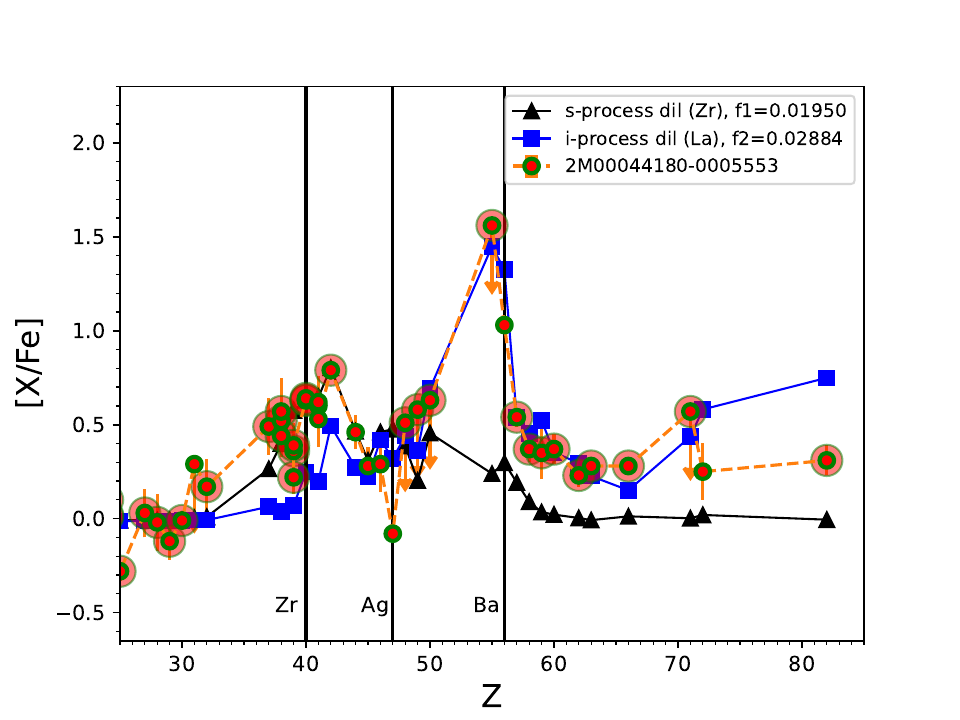}
  \caption{Elemental abundances beyond Fe of the four stars compared with the abundance pattern of case 3, that is, the s-process component peaked at the neutron-shell closure N=50 (with the observed Zr abundance as a reference to derive the dilution factor f1, black-triangles line) and the i-process component peaked at N=82 (with La as a reference to derive f2, calculated using seeds from the s-process component). The elements with the most reliable observational results (measurements or upper limits) are highlighted with large orange circles. 
  }
  \label{fig: scenario3_all}
\end{figure*}

\end{appendix}

\end{document}